# GRACIDIT: Graph–Circuit Digital Twin for Configuration-Induced Routing Delay Prediction in Zynq UltraScale+ FPGAs

Mostafa Darvishi, *Senior Member, IEEE*

***Abstract*— Configuration-induced perturbations in SRAM-based FPGAs may activate dormant programmable routing branches and increase path delay without immediately producing a functional error. Although prior studies have separately investigated the electrical origin of these delay changes, their in-situ detection, and the topology of commercial routing fabrics, a scalable method for predicting their timing impact at programmable interconnection point and routed net granularity remains unavailable. This paper presents GRACIDIT, a graph–circuit digital twin framework for predicting configuration-induced routing-delay degradation in Zynq UltraScale+ FPGAs. The proposed framework extracts the routing-resource graph of the XCZU7EV programmable fabric from the vendor design database, identifies inactive PIPs adjacent to active routes, and represents each candidate perturbation through its branch topology, geometric span, fan-out, physical region, and downstream loading. These graph features are combined with a calibrated reduced-order electrical model to estimate the delay introduced by single and cumulative routing-branch activations. Controlled configuration-equivalent perturbations are generated on a ZCU104 platform and characterized using complementary routing-dominated ring oscillators and phase-swept path monitors. The resulting model associates predicted delay shifts with available timing slack to rank vulnerable PIPs and routed nets and to construct a spatial vulnerability atlas of the programmable fabric. Experimental evaluation demonstrates a mean absolute prediction error of 7.8 ps, achieves 87.4% recall for slack-violating perturbations, and attains a Recall@10 of 0.90 for the most vulnerable routing resources. The proposed methodology provides a practical foundation for vulnerability-aware routing, optimized monitor placement, targeted configuration scrubbing, and adaptive route recovery in reliability-critical FPGA systems.**

## I. INTRODUCTION

Field-programmable gate arrays (FPGAs) have evolved from relatively simple programmable-logic devices into heterogeneous computing platforms integrating large reconfigurable fabrics, embedded processors, memories, arithmetic engines, high-speed interfaces, and system-level interconnects. Their ability to specialize the hardware datapath after fabrication makes them attractive for applications requiring high performance, low latency, adaptability, or rapid deployment. This flexibility, however, is enabled by an extensive programmable-routing network whose switches, multiplexers, and shared wires impose substantial area and timing costs compared with fixed interconnects. As transistor scaling has progressed, wire resistance, parasitic loading, and routing-resource organization have become increasingly important determinants of FPGA performance, making programmable-interconnect characterization and optimization central to modern FPGA architecture and CAD research [1], [2], [3].

In SRAM-based FPGAs, the selected logic functions and routing connections are maintained by volatile configuration-memory cells [4]. An upset affecting these cells can modify a look-up-table function, storage configuration, routing multiplexer, or programmable interconnect point (PIP) [5]. This susceptibility is particularly important in radiation-exposed and reliability-critical applications, where configuration-memory errors may persist until the affected frame is rewritten [6], [7], [8]. Conventional reliability techniques—including configuration scrubbing, modular redundancy, fault injection, and reliability-aware physical design—have therefore primarily focused on identifying configuration bits that produce observable functional failures and reducing their system-level impact [9], [10], [11].

A routing-related configuration upset does not, however, always produce an immediate Boolean failure. Circuit-level investigations have demonstrated that an unintentionally activated programmable switch may connect an unused routing segment to a live signal path [12]. The added branch introduces parasitic capacitance and possibly additional switch and downstream loading, increasing the signal-propagation delay while leaving the logical connectivity of the intended path apparently unchanged [5]. Subsequent work introduced an in-situ monitor capable of detecting single and cumulative routing-delay changes and experimentally associated these changes with configuration-memory locations [6]. Neutron-irradiation experiments further demonstrated automated traversal of switch matrices, controlled construction of routing-dominated oscillators, and measurement of induced delay changes with picosecond-scale resolution [7]. These results establish configuration-induced delay degradation as a physically measurable failure precursor rather than merely a theoretical consequence of routing faults.

Despite these advances, an important design-time question remains unresolved: *Given an arbitrary routed net, which*

Mostafa Darvishi is with the Department of Electrical Engineering, École de technologie supérieure (ÉTS), Montreal, Canada. (e-mail: darvishi@ieee.org).

*inactive PIPs can become harmful parasitic attachments, how much delay would each attachment introduce, and whether that delay would exhaust the net's available timing slack?* Existing delay-monitoring methods determine that a delay shift has occurred after implementation or during operation, whereas detailed circuit simulations require internal electrical assumptions and become impractical when independently applied to the large number of routing resources surrounding a complete design. Similarly, conventional static timing analysis (STA) evaluates the nominal configured route under specified timing corners, but it does not normally enumerate adjacent inactive PIPs or predict the incremental delay resulting from their unintended activation. Consequently, two routed nets having comparable nominal delay and slack may exhibit substantially different configuration-induced timing vulnerability.

Recent routing-architecture studies provide an important foundation for addressing this limitation. *NetCracker* models a commercial Xilinx 7-Series routing fabric as a directed graph whose vertices represent PIP junctions and whose edges represent their internal and external connectivity. Its analyses expose routing-channel composition, switch-box adjacency, physical direction, fan-in, fan-out, and spatial architectural diversity using information available through vendor design tools [8]. Nevertheless, such graph-level architectural descriptions do not assign an electrical delay consequence to the activation of a specific dormant branch. Conversely, circuit models describe the electrical origin of the additional delay but do not provide a scalable mechanism for applying that model to every relevant PIP, routed node, and physical region of a modern device. This separation between architectural topology and electrical behavior prevents existing methods from producing a design-specific, PIP-level map of routing-delay vulnerability.

The need to bridge this separation becomes more pronounced in advanced FPGA technologies. Modern programmable-interconnect modeling shows that scaled wires, routing multiplexers, shared-wire loading, and physical design constraints interact in ways that cannot be represented reliably by simply transferring parameters or architectural assumptions from older device families [3]. Furthermore, commercial fabrics are spatially heterogeneous: switch-box structure and connectivity may vary near logic, memory, arithmetic, clocking, interface, and device-edge regions [8]. A delay-prediction method intended for a Zynq UltraScale+ device must therefore derive its routing taxonomy from the target fabric and calibrate its effective electrical behavior on that device rather than inheriting 7-Series or Virtex-era resource classes and circuit parameters.

Reliability-oriented placement and routing algorithms have previously reduced the exposure of sensitive signals to configuration upsets by modifying resource selection, physical separation, or routing cost [9], [10]. More recently, bitstream-level design-space exploration has enabled automated evaluation of cost–reliability trade-offs for partially redundant FPGA implementations [11]. These approaches provide valuable mechanisms for fault avoidance and mitigation, but their optimization targets are generally based on functional criticality, bridging-fault likelihood, redundancy coverage, or system-level error response. They do not predict the continuous delay increment associated with each candidate routing attachment, nor do they combine that increment with path slack to determine the probability of a timing violation. Thus, a routed resource may be treated as critical because it can alter functionality, or noncritical because no immediate logical error is observed, while its potential to cause latent timing degradation remains unquantified.

This paper addresses this gap by introducing a graph–circuit digital twin for configuration-induced routing-delay prediction in Zynq UltraScale+ FPGAs. The proposed framework extracts an UltraScale+-native graph of active routes and neighboring inactive PIPs from the XCZU7EV device database. Each candidate routing perturbation is represented using its attachment location, branch topology, geometric displacement, routing-resource class, fan-out, downstream connectivity, and surrounding physical region. These graph-derived features are combined with a calibrated reduced-order electrical model to predict the incremental delay produced by single and cumulative routing-branch activations. Controlled *configuration-equivalent perturbations* on an AMD/Xilinx ZCU104 platform provide deterministic reference and perturbed implementations, while routing-dominated ring oscillators (ROs) and phase-swept path monitors provide complementary hardware measurements for model calibration and validation. Here, a *configuration-equivalent perturbation* denotes a deterministic reference–variant pair that preserves the intended active route and logical function while electrically attaching a specified dormant routing component. It reproduces the loading consequence of the corresponding configuration-state change without implying that the controlled experiment itself is radiation-induced.

Fig. 1 summarizes the proposed workflow. Conventional STA supplies the nominal route delay $d_{\text{nom}}$ and slack $S$, while the extracted UltraScale+ routing graph identifies inactive PIPs capable of attaching dormant components to the active route. The graph–circuit model combines branch topology, loading, physical context, and calibrated electrical parameters to predict the delay increment $\Delta d$, the perturbed path slack $S_{\text{pert}} = S - \Delta d$, uncertainty, and slack-exhaustion probability. Controlled route perturbations, routing-dominated ROs, and phase-swept path monitors on the ZCU104 platform provide model calibration and validation. The principal contributions are:

1. an UltraScale+-native extraction flow that constructs the XCZU7EV routing graph and enumerates dormant branches adjacent to implemented routes;
2. a physics-informed graph–circuit model that predicts delay increments for single and cumulative branch activations;
3. a controlled configuration-equivalent validation methodology linking selected PIPs to configuration-frame changes;
4. dual-channel calibration using routing-dominated

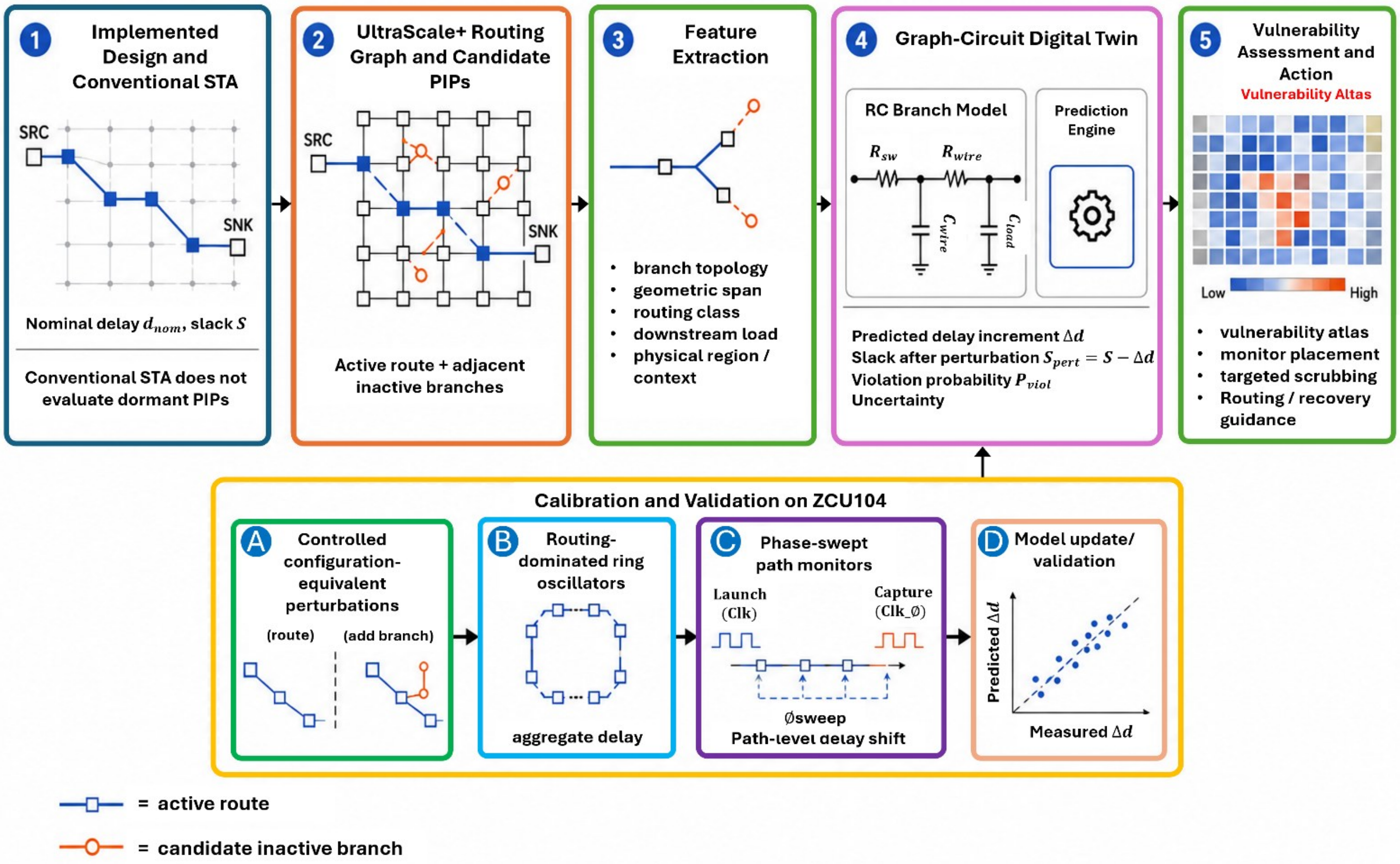

Fig. 1. Overview of the proposed graph–circuit digital twin framework for configuration-induced routing-delay prediction in Zynq UltraScale+ FPGAs.

oscillators and phase-swept path monitors; and

5. PIP-, net-, frame-, and region-level vulnerability assessment combining predicted delay, path slack, uncertainty, and, when available, perturbation-occurrence information.

The remainder of this paper is organized as follows. Section II reviews configuration-induced routing-delay degradation, timing-observation techniques, commercial routing-architecture extraction, and reliability-aware FPGA CAD. Section III formulates the UltraScale+ routing graph, configuration-equivalent perturbations, and timing-failure conditions. Section IV presents the proposed graph–circuit digital twin, while Section V describes the ZCU104 implementation and experimental methodology. Section VI reports and discusses the experimental results, and Section VII concludes the paper and outlines future work.

## II. BACKGROUND AND RELATED WORK

### A. Configuration-Memory Upsets and Routing-Delay Degradation

The programmable logic (PL) and interconnection network of SRAM-based FPGAs are controlled by configuration-memory cells that determine logic functions, multiplexer selections, and programmable routing connections. A perturbation of a routing-related configuration bit may disable an intended PIP, activate an unintended PIP, or modify the connectivity of a routing multiplexer. The first two consequences are commonly modeled as open and short faults, respectively. However, an unintentionally enabled PIP does not necessarily connect two actively driven signals or immediately change the Boolean behavior of the implemented circuit. It may instead attach an otherwise unused wire segment to a routed signal, thereby creating a persistent parasitic branch [5], [12].

Circuit-level modeling has shown that the dominant delay contribution in this case arises from the pass transistor associated with the activated PIP and the capacitance of the unintentionally connected interconnect segment. The added branch increases the effective load observed by the upstream driver and can delay both rising and falling transitions propagating along the original route [5]. The magnitude of the resulting delay change depends on the electrical characteristics of the activated switch, the length and topology of the attached wire, the downstream resources reachable through that branch, and the electrical state of the affected path. Simulated delay changes obtained from pass-transistor and interconnect models were shown to agree closely with proton-irradiation measurements, establishing a physical link between configuration-state perturbations and additional combinational delay.

A single activated branch produces a single delay change, whereas multiple perturbations may create cumulative delay degradation. Successive PIP activations can either connect several branches directly to the active route or extend an already attached parasitic network through additional switches and wires. The latter mechanism is particularly important because a first perturbation may expose new downstream PIPs that were not electrically connected to the routed signal before the initial event. Circuit simulations have demonstrated that these progressively connected branches can produce substantially larger delay changes than an isolated attachment [5], [7].

Hardware studies subsequently confirmed both single and cumulative routing-delay changes. The in-situ monitor in [6] reported single delay increments of 29–151 ps and cumulative increments of 279–309 ps, while associating observed events with configuration-memory locations. Routing-dominated oscillators implemented on a Zynq-7000 device later resolved radiation-induced changes of approximately 5 ps and demonstrated dependence on routing-resource class and physical implementation [7]. These approaches establish physical observability but remain limited to selected monitored nodes or purpose-built oscillators rather than arbitrary routed paths.

Fault emulation and controlled configuration perturbation provide repeatable alternatives to beam testing, but the relationship between an emulated configuration change and the physical fault mechanism must be validated carefully. The work presented in [13] showed that fault-emulation methodologies for SRAM-based FPGAs require explicit validation of fault locations, configuration effects, and observed failure modes before their results can be used as substitutes for radiation data. Accordingly, the present work employs controlled configuration-equivalent perturbations to reproduce the electrical consequence of an additional routing attachment. It does not claim that each controlled experiment constitutes a radiation-induced event; rather, the prior circuit-level and irradiation studies establish the physical relevance of the branch-activation mechanism being modeled.

### *B. In-Situ Timing Observation and Timing-Error Resilience*

In-situ timing-error detection has been studied extensively as a means of reducing conservative timing margins and maintaining correct operation under process, voltage, temperature, and aging variability. *Razor* technique augments selected state elements with delayed sampling to identify late-arriving transitions and initiate architectural recovery [14]. TIMBER uses time borrowing and error relaying across successive pipeline stages to mask timing errors without imposing the full nominal timing margin at every boundary [15]. The time-dilation technique similarly detects a timing violation and temporarily extends the available execution interval to prevent incorrect state capture [16]. These approaches demonstrate the value of observing timing behavior during circuit operation, but they primarily detect or tolerate a timing error after a transition approaches or crosses a sampling boundary. *They do not determine which dormant FPGA routing resource caused the delay*, *quantify the delay associated with every candidate PIP*, *or rank vulnerable routes before deployment*.

FPGA-specific delay monitors face additional implementation constraints because the observation logic itself must be placed and routed through the same programmable fabric as the monitored circuit. The monitor in [6] was designed to detect routing-delay changes at selected sensitive nodes while imposing substantially less overhead than inverter-chain- or scan-based alternatives. Its adjustable threshold permits the timing budget of a monitored node to be considered, but its loading and detection resolution vary with the monitor's physical distance from the signal under test. The reported implementation added approximately 31–40 ps of net delay when placed close to the monitored node and substantially more when placed farther away. Consequently, instantiating a monitor at every potentially vulnerable node would introduce both resource and timing overhead, motivating selective monitor placement rather than exhaustive instrumentation.

Routing-dominated ROs provide a complementary measurement mechanism. A perturbation that changes one or more constituent routing delays shifts the oscillator period and can therefore be detected through frequency measurement. Their repeated transitions and long accumulated routes offer high sensitivity to picosecond-scale changes, but the measured frequency shift represents an aggregate variation over the complete oscillation loop. *It does not by itself localize the affected path segment or determine whether an equivalent delay would violate the slack of a synchronous user path* [7].

In the proposed framework, ROs provide sensitive aggregate measurements for electrical calibration, whereas phase-swept monitors provide path-local timing displacement. Both channels calibrate a predeployment predictor rather than serving as the final vulnerability indicator.

### *C. Routing-Architecture Representation and Commercial-FPGA CAD*

An FPGA routing fabric can be represented as a directed graph in which vertices denote routing junctions or wire endpoints and edges denote fixed or programmable connections. Such a representation supports traversal, reachability analysis, fan-in and fan-out extraction, routing-resource classification, and path-cost computation. *NetCracker* formalized this approach for Xilinx 7-Series devices by representing switch-box PIP junctions and their connections in a vendor-independent graph [8]. Its analysis passes classify PIP-junction roles, infer routing-wire direction vectors, calculate routing-channel composition, derive adjacency relationships, and identify switch-box diversity. The *NetCracker* results also show that a single uniform switch-box abstraction is insufficient for a commercial device. Although many switch boxes share similar external connectivity, deviations occur near hardened resources, specialized columns, and device boundaries. In addition, vendor wire names do not always uniquely express geometric displacement or secondary destinations, motivating explicit coordinate-based descriptions of routing edges [8]. These observations are directly relevant to delay-vulnerability prediction because the loading introduced by an activated branch can depend not only on its reported resource name but also on its exact span, direction, reachable nodes, and surrounding tile context.

Commercial-device CAD frameworks provide mechanisms

Table I. Comparison of prior FPGA routing-reliability approaches with the proposed graph–circuit digital twin

| Approach | Commercial PIP-Level Topology | Selected PIP-to-Configuration Association | Quantitative Perturbation-Delay Prediction | Single and Cumulative Perturbations | Hardware or Radiation Validation | Path-Slack Awareness | Predeployment Vulnerability Output |
|---|---|---|---|---|---|---|---|
| **Circuit-level routing-delay modeling [5]** | ◐ | — | ✓ | ✓ | ✓ | — | — |
| **In-situ routing-delay monitor [6]** | ◐ | ✓ | — | ✓ | ✓ | ✓ | — |
| **Routing-resource radiation characterization [7]** | ◐ | — | — | ◐ | ✓ | — | — |
| **NetCracker routing analysis [8]** | ✓ | — | — | — | — | — | — |
| **Reliability-aware FPGA CAD [9], [10], [19]** | ◐ | ◐ | — | — | ◐ | ◐ | ◐ |
| **RapidWright and RWRoute [17], [18]** | ✓ | — | — | — | ◐ | ✓ | — |
| **Proposed Work** | ✓ | ✓ | ✓ | ✓ | ✓ | ✓ | ✓ |

✓ : explicitly supported; ◐ : partially or indirectly supported; — : not addressed.

for accessing and manipulating such low-level routing objects. *RapidWright* exposes design checkpoints and device resources for customized physical implementation on AMD/Xilinx FPGAs [17]. *RWRoute* extends this environment with an open-source timing-driven router for commercial devices and uses device-specific connectivity and timing information to guide route construction [18]. These tools demonstrate that academic algorithms can operate directly on contemporary commercial routing databases rather than being restricted to simplified island-style fabrics.

Existing commercial-device frameworks provide graph connectivity and nominal timing information, but they do not evaluate the electrical consequence of activating adjacent inactive PIPs. Because resistance, switch loading, shared-wire capacitance, and spatial heterogeneity vary across technologies and device regions [3], the proposed method combines the graph extracted from the XCZU7EV database with effective electrical parameters calibrated on the same device.

### *D. Reliability-Aware FPGA Physical Design*

Reliability-aware CAD techniques attempt to reduce the probability that configuration-memory faults affect an implemented design. The work presented in [9] incorporated reliability considerations into placement and routing to reduce the sensitivity of SRAM-based FPGA implementations. The SEU awareness into FPGA routing, considering the likelihood and impact of routing-related bridging faults when selecting resources was reported in [10]. Other CAD-based approaches combine redundancy with placement and routing objectives or exploit unused resources to reduce the predicted system failure rate [19]. More recently, bitstream-level design-space exploration has been used to evaluate cost–reliability trade-offs among alternative partially redundant implementations [11]. The CAD-based mitigation study in [19], for example, reported that adding SEU-aware objectives to packing, placement, and routing could lower estimated failure rates without requiring complete circuit replication.

These methods establish that routing decisions materially influence FPGA reliability. Their optimization targets, however, are generally based on the number of sensitive configuration bits, estimated functional failure probability, bridging-fault exposure, physical separation, or redundancy coverage. Such metrics are appropriate when an upset is classified according to whether it changes circuit functionality. *They do not distinguish between two nonfunctional routing perturbations that introduce, for example, a few picoseconds and several hundred picoseconds of additional delay*.

A complete timing-vulnerability assessment requires three quantities that are not jointly available in prior works: (i) the set of inactive routing branches capable of attaching to a particular active route, (ii) the delay distribution associated with activating each branch, and (iii) the available slack of the affected synchronous path. Without this combination, a structurally exposed net may be classified as highly vulnerable even when the induced delay is far below its slack, whereas a low-probability perturbation on a near-critical path may be underestimated despite its ability to cause immediate timing failure.

Table I compares prior approaches with the capabilities required for design-time routing-delay-vulnerability prediction. Existing methods provide individual elements of the solution, but none jointly combines commercial PIP-level topology, selected configuration association, calibrated delay prediction, cumulative perturbation analysis, path slack, and predeployment vulnerability assessment. The proposed method is distinguished by the quantitative connection it establishes among routing topology, electrical loading, configuration state, and sink-specific timing margin.

## III. ULTRASCALE+ ROUTING AND CONFIGURATION PERTURBATION MODEL

This section establishes the architectural and mathematical representation used by the proposed digital twin. The formulation separates three quantities that are commonly treated independently: (i) the physical route implemented by the vendor CAD flow, (ii) the inactive programmable resources that can become electrically attached to that route, and (iii) the path-specific timing margin available to tolerate the resulting delay. The corresponding prediction and calibration procedures are developed in Section IV.

### *A. Target Fabric and Physical-Routing Objects*

The target platform is the AMD/Xilinx ZCU104 evaluation board containing an XCZU7EV Zynq UltraScale+ MPSoC. The scope of this work is restricted to the PL interconnection network of the fabric; the processing system is employed only for experiment control and data acquisition. In the UltraScale architecture, CLBs are arranged in a regular array and access horizontal and vertical general-routing resources through associated switch matrices. Comparable switch-matrix interfaces connect other programmable-logic resources, including DSP and block-memory columns, resulting in a physically heterogeneous routing environment [20], [21].

A routed design checkpoint is represented using the physical objects exposed by the AMD/Xilinx Vivado device database. These include nets, site pins, nodes, wires, tiles, PIPs, and clock regions. In particular, the Vivado Tcl interface permits PIPs to be queried from nets, nodes, wires, tiles, or other PIPs and supports both uphill and downhill graph traversal [22]. The extraction flow therefore does not infer connectivity from graphical device views or naming conventions alone; it obtains the physical-resource relationships directly from the target-device database. Let the device-routing fabric be represented by the directed attributed

$$\mathcal{G}_D = (\mathcal{V}, \mathcal{E}_P, \mathcal{E}_F) \quad (1)$$

graph of Equation (1),

where $\mathcal{V}$ contains routing nodes and terminal vertices, $\mathcal{E}_\mathrm{P}$ is the set of programmable edges corresponding to PIPs, and $\mathcal{E}_\mathrm{F}$ contains fixed connections between routing wires, site pins, and resource terminals. The direction assigned to a programmable edge follows the direction reported by the device database. Bidirectional or structurally asymmetric resources are retained according to their exposed connectivity rather than being replaced by an idealized switch-box model. Each routing edge $\boldsymbol{e}$ is associated with an attribute vector of

$$a_e = [\tau_e, x_e, y_e, r_e, \Delta x_e, \Delta y_e, d_e^-, d_e^+, \lambda_e, \chi_e] \quad (2)$$

Equation (2),

where:

- $\tau_e$ denotes the vendor-reported resource or PIP type;
- $(x_e, y_e)$ denotes its physical tile location;
- $r_e$ identifies its clock region;
- $(\Delta x_e, \Delta y_e)$ describes the geometric displacement between its endpoint tiles;
- $d_e^-$ and $d_e^+$ denote the local fan-in and fan-out degrees;
- $\lambda_e$ represents the routing-resource span or branch-depth descriptor; and
- $\boldsymbol{\chi_e}$ identifies the surrounding physical context, such as CLB-, BRAM-, DSP-, clocking-, or device-boundary adjacency.

The exact properties available for a particular object are obtained using list_property and report_property in the locked Vivado IDE release. Consequently, Equation (2) defines a common architectural representation, while the final feature set retains only quantities that can be extracted reproducibly from the XCZU7EV database.

### *B. UltraScale+-Native Routing Taxonomy*

The routing-resource classes used in this study are not transferred from Virtex-5 or Xilinx 7-Series devices. The earlier routing taxonomy distinguished resources such as single-, double-, quad-, long-, bounce-, pin-feed-, and outbound interconnects and demonstrated that these classes can exhibit different radiation-induced delay behavior [7]. *NetCracker* similarly showed that commercial 7-Series routing wires and PIP junctions can be classified according to direction, span, connectivity, and switch-box role [8]. However, UltraScale devices differ from earlier generations in CLB organization and physical-routing structure, and their switch matrices connect a broader set of heterogeneous resources [21].

The proposed taxonomy is therefore derived from the extracted graph itself. Two routing edges $e_i$ and $e_j$ are assigned to the same structural class only when they share the

$$C(e) = (\tau_{src}, \tau_{dst}, \Delta x, \Delta y, q_{PIP}, q_{term}, \chi) \quad (3)$$

normalized descriptor of Equation (3),

where $\tau_\mathrm{src}$ and $\tau_\mathrm{dst}$ are the endpoint tile or node classes, $q_\mathrm{PIP}$ is the PIP-junction category, $q_\mathrm{term}$ denotes the terminal-resource class, and $\chi$ represents the physical-region context. Vendor resource names are preserved as identifiers but are not used as the sole basis of classification because a name may not fully encode geometric span, secondary destinations, or spatial variation [8]. The extracted routing resources are organized hierarchically into four levels:

1. **Local junction class:** connectivity internal to, or immediately adjacent to, a resource tile;
2. **Geometric wire class:** direction and displacement of the inter-tile connection;
3. **Terminal class:** the type of source or destination resource reached by the branch;
4. **Spatial-context class:** the physical region in which an otherwise similar structure occurs.

This hierarchy permits the predictor to determine whether nominally similar branches exhibit consistent behavior across the device or require region-specific correction. It also avoids over-fragmenting the dataset into thousands of vendor-specific wire names that may each have too few measurements for reliable calibration.

### *C. Active Routes and Sink-Specific Path Representation*

A placed-and-routed net is generally a rooted routing tree rather than a single linear path. Vivado represents fixed routes using directed routing strings in which nested branches denote fan-out from a common routing trunk [23]. For a net $n$ with driver $u_n$ and sink set $\mathcal{S}_n$, its active routing subgraph is

$$\mathcal{R}_n = (\mathcal{V}_n, \mathcal{E}_n^A) \subseteq \mathcal{G}_D \quad (4)$$

defined as Equation (4),

where $\boldsymbol{\mathcal{E}_n^A}$ contains the fixed and programmable resources used by the implemented route. Directed-route constraints and fixed-route properties allow selected routes to be preserved when controlled route variants are generated [23]. Because different sinks of the same net may share only part of the routing tree, perturbation analysis is performed at sink granularity. The route from the driver of $n$ to sink $s \in \mathcal{S}_n$ is

$$\mathcal{R}_{n,s} = \left(\mathcal{V}_{n,s}, \mathcal{E}_{n,s}^A\right) \quad (7)$$

written as Equation (7),

where $\mathcal{E}_{n,s}^{\mathrm{A}} \subseteq \mathcal{E}_n^{\mathrm{A}}$. A perturbation attached to the shared routing trunk may affect several sinks, whereas one attached after a branch point affects only the corresponding downstream sink. This distinction is essential because the same activated PIP can have different timing significance depending on the sink path and its available slack. For each sink-specific route, the extraction process records:

- its ordered active PIPs and nodes;
- branch points shared with other sinks;
- source and destination site pins;
- post-route interconnect and logic delay;
- setup and hold slack;
- clock domain and clock region;
- physical route length and resource composition; and
- the inactive PIPs incident on each active routing node.

### *D. Dormant-Branch Candidate Enumeration*

A *dormant branch candidate* is an inactive programmable connection whose activation electrically attaches an otherwise unused routing component to an active signal path while preserving the intended source-to-sink connection. For a sink-specific route $\mathcal{R}_{n,s}$, the first-order candidate set is denoted as Equation (5),

$$\mathcal{B}_{n,s}^{(1)} = \left\{p \middle| \in \middle| \mathcal{E}_P \setminus \mathcal{E}_n^A \;\; |\partial p \cap \mathcal{V}_{n,s}| = 1,\, drv\left(\mathcal{H}_p\right) = 0\right\} \quad (5)$$

where $\partial p$ denotes the two routing-node endpoints of PIP $p$, and $\mathcal{H}_p$ is the inactive routing component reached through the endpoint outside the active path. The condition $\mathrm{drv}\left(\mathcal{H}_p\right) = 0$ excludes components containing another active driver.

Both uphill and downhill adjacency are examined because the electrical attachment of interest is defined by incidence on the active route, whereas the database direction indicates the legal routing direction of the PIP. Uphill and downhill queries are used only to discover PIPs incident on the active routing node. Candidate retention is determined by electrical visibility rather than database direction alone. Passive or bidirectional switch resources are retained when enabling the PIP exposes a nonzero branch impedance to the active route. Directionally buffered resources are rejected when their internal direction prevents the attached branch from loading the analyzed signal path. Thus, graph adjacency alone is not sufficient for candidate acceptance. Vivado IDE supports both forms of traversal through the get_pips interface [22]. For each candidate $b \in \mathcal{B}_{n,s}^{(1)}$, the inactive component is expanded from the dormant endpoint to form the candidate branch graph

$$\mathcal{H}_b = (\mathcal{V}_b, \mathcal{E}_b) \quad (6)$$

denoted in Equation (6).

For each candidate $b$, the dormant component $\mathcal{H}_b$ is extracted by a deterministic breadth-first traversal beginning at the endpoint outside the active route. The traversal follows fixed routing edges and programmable edges that are already enabled in the reference configuration; an inactive PIP is crossed only when it belongs explicitly to the perturbation set $B$. Expansion terminates upon reaching a resource terminal, an active routing resource, a second active driver, a predefined isolation boundary, or the maximum graph depth $D_{\max}$. The value $D_{\max} = 8$ routing hops is fixed before model calibration. A sensitivity analysis using $D_{\max} + 4 = 12$ hops confirms that increasing the limit changes the retained branch descriptors by less than 1.8%. In this context, a configured downstream resource denotes a routing or terminal resource whose controlling configuration state is already active but that remains electrically disconnected from the analyzed net until the candidate PIP is enabled. The depth limit is increased until the extracted structural descriptors stabilize or the reachable component is fully enumerated. A valid delay-only candidate

$$\mathcal{E}_{n,s}^A \subseteq \mathcal{E}_{n,s}^A \cup \mathcal{E}_b, \quad (8)$$

$$drv(\mathcal{H}_b) = 0,$$

$$F_{logic}^{(b)} = F_{logic}^{(0)},$$

$$\Delta d_b \neq 0,$$

must satisfy four conditions of Equation (8),

where $F_{\mathrm{logic}}^{(0)}$ and $F_{\mathrm{logic}}^{(b)}$ are the logical functions of the reference and perturbed implementations, respectively. Thus, the intended route is retained, no actively driven net is shorted to it, and the candidate is evaluated as a timing perturbation rather than as a Boolean routing fault. This exclusion is deliberate. Open PIPs, shorts between two driven nets, logic-function corruption, and state-element upsets remain important FPGA fault mechanisms, but they do not belong to the delay-only prediction problem addressed here. The underlying physical motivation is the parasitic-branch mechanism previously established through circuit simulation, controlled emulation, and irradiation [5]– [7].

### *E. Single and Cumulative Perturbation Classes*

The candidate branches are divided into four perturbation classes according to the topology introduced by their activation:

#### 1) Single Leaf Attachment

A single PIP connects one inactive wire segment to the active route without reaching an additional logic input or routing load. The perturbation set contains one programmable edge as denoted in Equation (9),

$$\mathcal{P}_b = \{p_1\} \tag{9}$$

and the branch is dominated by the switch and wire loading.

#### 2) Single Loaded Attachment

A single activated PIP exposes a branch containing one or more configured downstream junctions or terminal loads. Although only one new PIP is activated, the effective branch capacitance can be greater than that of an isolated wire because previously configured downstream resources become electrically connected.

#### 3) Parallel Cumulative Attachment

Two or more independent branches are attached to different nodes, or to the same node, of the active route of Equation (10).

$$\mathcal{P}_B = \{p_1, p_2, \cdots, p_K\}, \quad K > 1 \tag{10}$$

The branches load the active route in parallel but may affect different portions of its distributed resistance–capacitance structure.

#### 4) Serially Extended Cumulative Attachment

A first perturbation connects a dormant branch to the active route, and one or more additional PIP activations extend that branch into a larger inactive network. The latter case reproduces the cumulative mechanism in which an initial configuration change makes further downstream resources electrically relevant [5], [6]. For a cumulative perturbation set $B$, the complete attached component is denoted as Equation (11).

$$\mathcal{H}_B = \bigcup_{b \in B} \mathcal{H}_b \tag{11}$$

No assumption is made that the cumulative delay is the arithmetic sum of the individual delay changes, as denoted in

$$\Delta d_B \neq \sum_{b \in B} \Delta\, d_b \quad \text{in general} \tag{12}$$

Equation (12),

This is because branch interactions depend on attachment locations, shared wire segments, distributed resistance, transition slew, and the topology of the combined inactive network. Later in Section IV, the graph–circuit model therefore will evaluate $\mathcal{H}_B$ as a unified perturbation structure rather than merely summing independent predictions.

### *F. Configuration-State and Frame Association*

Let the relevant configuration state of the device be

$$\mathbf{z} = [z_1, z_2, \cdots, z_M]^T \tag{13}$$

represented by the binary vector of Equation (13),

where $z_i$ denotes a configuration-memory value included in the analyzed frame set. The reference implementation has state $z^{(0)}$, while the controlled variant associated with perturbation $b$has state $\mathbf{z}^{(b)}$. The differential configuration set is denoted in

$$\mathcal{D}_b = \left\{ i \middle| \; z_i^{(0)} \oplus z_i^{(b)} = 1 \right\} \tag{14}$$

Equation (14).

A one-to-one correspondence between a PIP and a single configuration bit is not assumed. Depending on the routing multiplexer and encoding, the state transition associated with a candidate PIP can involve one or more differing configuration locations. The candidate mapping is therefore defined as

$$\mathcal{M}(b) = \{(F, w, k)\}_b \tag{15}$$

Equation (15),

where $F$ is the configuration-frame address, $w$ is the word index within the frame, and $k$ is the bit index within that word. For UltraScale+ devices, the *Frame Address Register* is organized into block-type, row, column, and minor-address fields, and each configuration frame contains 93 32-bit words [24]. These documented fields provide the frame-level organization used by the mapping procedure, but they do not by themselves identify which PIP is controlled by a particular word and bit.

Selected PIP-to-configuration associations are consequently derived through controlled differential implementations. The placement, active route, logic configuration, clocking, and unrelated routing are fixed, while the target branch state is changed. A candidate association is accepted only when:

1. the same differential location appears across repeated paired builds;
2. reversing the target route change restores the corresponding configuration state;
3. no unrelated active resource changes in the locked design region;
4. frame readback agrees with the expected differential state; and
5. the observed routing and delay behavior is consistent with the selected branch.

Differential bitstream analysis has previously been used to infer mappings between FPGA programmable resources and undocumented configuration locations [25]. In this paper, however, mapping is restricted to the PIPs required for model calibration and validation; *complete reverse engineering of the XCZU7EV bitstream is neither assumed nor required*.

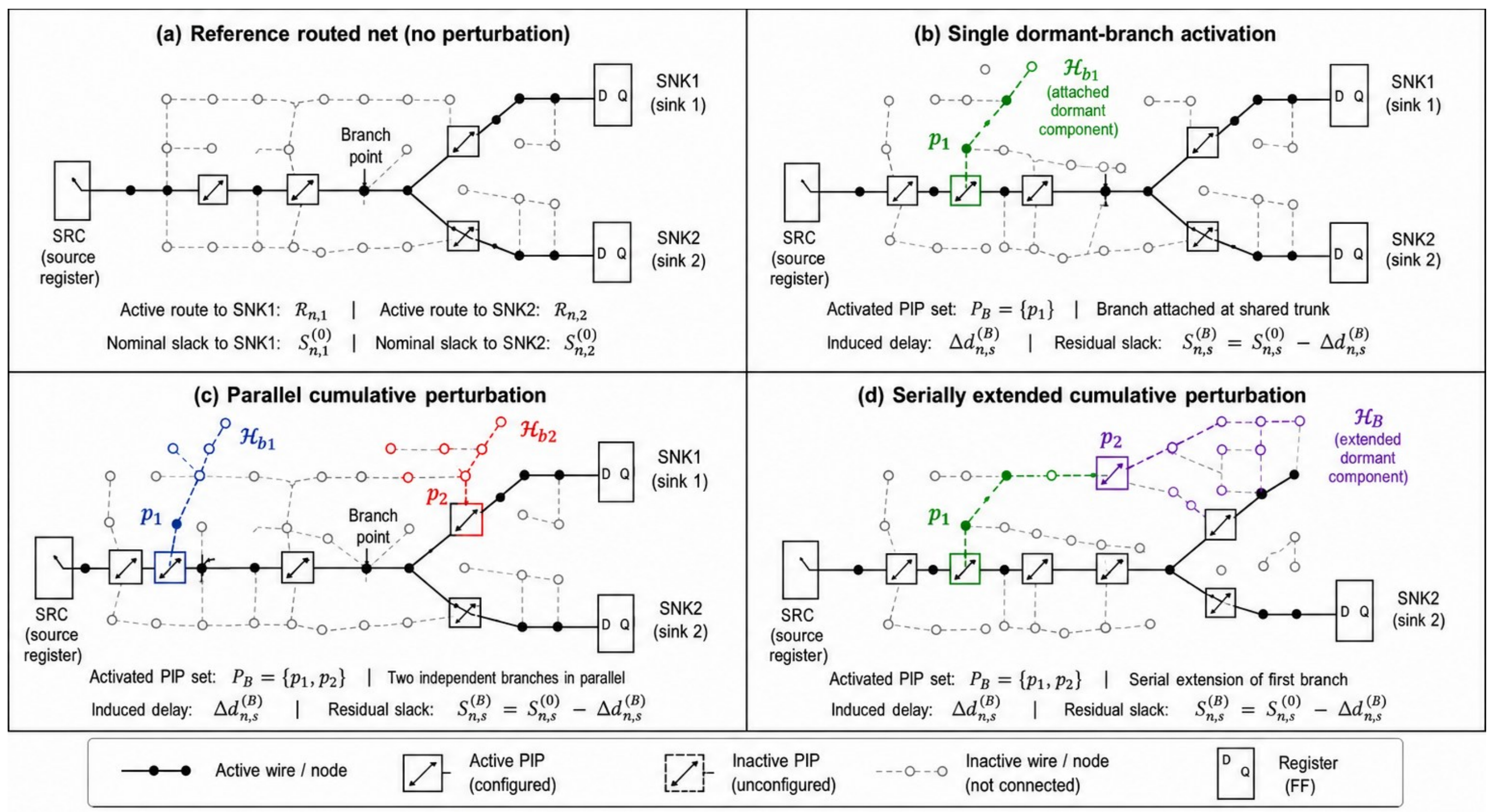

Fig. 2. Routing-perturbation classes considered in the proposed model. (a) Reference routed net with a shared trunk, fan-out point, and two sink-specific paths. Solid black resources form the configured route, whereas dashed gray resources are dormant. (b) Single dormant-branch activation through $p_1$, attaching $\mathcal{H}_{b1}$ to the shared trunk. (c) Parallel cumulative perturbation produced by independent activation of $p_1$ and $p_2$, with shared-trunk and sink-local effects. (d) Serially extended cumulative perturbation in which $p_2$ enlarges the component initially attached through $p_1$. In all modeled cases, the intended route and logical function are preserved, while the induced delay $\Delta d_{n,s}^{(B)}$ reduces the sink-specific slack from $S_{n,s}^{(0)}$ to $S_{n,s}^{(B)}$.

### G. Timing-Failure Condition

Let $d_{n,s}^{(0)}$ denote the post-route data-path delay associated with sink $s$of net $n$, including the relevant source-register, combinational, routing, and destination-register timing terms. Let $T_{n,s}^{\text{req}}$ be the required arrival time reported by STA. The nominal setup slack is written as Equation (16), which after activation of perturbation set $B$, the path delay will be as noted in Equation (17) with the residual slack on Equation (18).

$$S_{n,s}^{(0)} = T_{n,s}^{req} - d_{n,s}^{(0)} \tag{16}$$

$$d_{n,s}^{(B)} = d_{n,s}^{(0)} + \Delta d_{n,s}^{(B)} \tag{17}$$

$$S_{n,s}^{(B)} = S_{n,s}^{(0)} - \Delta d_{n,s}^{(B)} \tag{18}$$

A setup-timing violation occurs when $S_{n,s}^{(B)} < 0$, equivalently

$$\Delta d_{n,s}^{(B)} > S_{n,s}^{(0)} \tag{19}$$

when

Equation (19) demonstrates why perturbation severity cannot be determined from induced delay alone. For example, a 100 ps delay increase can be harmless on a path with 1 ns of slack but immediately critical on a path with 20 ps of slack. Conversely, ranking routes solely by nominal slack ignores substantial differences in the number and structure of dormant branches adjacent to them. Because the predicted delay has model and measurement uncertainty, the final risk quantity is

$$P_{\text{fail}}(n, s, B) = \Pr\left[\Delta d_{n,s}^{(B)} > S_{n,s}^{(0)} \mid x_B, \theta\right] \tag{20}$$

expressed probabilistically as Equation (20),

where $x_B$ is the graph-derived perturbation descriptor and $\theta$ contains the calibrated model parameters. The construction of $x_B$, estimation of $\theta$, and propagation of prediction uncertainty will be presented later in Section IV.

### H. Model Scope and Assumptions

The formulation is governed by the following scope boundaries:

***First***, the modeled perturbation is persistent until configuration correction or reconfiguration occurs; transient combinational pulses not associated with a lasting configuration-state change are outside the present model. ***Second***, the intended active route remains connected, and perturbations producing opens or contention between active drivers are excluded. ***Third***, configuration association is performed for selected PIPs rather than for every programmable resource in the device. ***Fourth***, the device database and extraction results are tied to the locked target part and Vivado IDE release because object properties and naming conventions can vary across device families and tool versions. ***Finally***, process, supply-voltage, temperature, placement, and measurement effects are not ignored; they enter the calibrated parameters and uncertainty model developed in the following section.

Fig. 2 defines the routing-perturbation classes considered in this work. Fig. 2(a) shows a reference routed net in which one source register drives two sink registers through a shared routing trunk and a fan-out point. The solid black wires and configured PIPs constitute the active routing tree, whereas the dashed gray wires, nodes, and unconfigured PIPs denote dormant routing resources that are not electrically connected to the implemented net. The sink-specific routes are denoted by $\mathcal{R}_{n,1}$ and $\mathcal{R}_{n,2}$, with corresponding nominal setup slacks $S_{n,1}^{(0)}$ and $S_{n,2}^{(0)}$.

Fig. 2(b) illustrates a single dormant-branch activation. Enabling $p_1$ attaches the previously disconnected component $\mathcal{H}_{b1}$ to the shared routing trunk without interrupting the intended paths or changing the implemented logical function. Because the attachment occurs before the fan-out point, the resulting electrical loading can influence the routes to both sinks. The induced delay $\Delta d_{n,s}^{(B)}$ reduces the residual slack of each affected sink path according to

$$S_{n,s}^{(B)} = S_{n,s}^{(0)} - \Delta d_{n,s}^{(B)}$$

Fig. 2(c) represents a parallel cumulative perturbation in which two independently activated PIPs, $p_1$ and $p_2$, attach the dormant components $\mathcal{H}_{b1}$ and $\mathcal{H}_{b2}$ at different locations along the active routing tree. The branch connected to the shared trunk may affect both sink paths, whereas the attachment placed after the fan-out point primarily affects the corresponding sink-local route only. This distinction motivates the sink-specific perturbation analysis introduced in Section III-C. Fig. 2(d) shows a serially extended cumulative perturbation. The first activated PIP, $p_1$, connects a dormant branch to the active route, while the subsequent activation of $p_2$ extends that branch into the larger component $\mathcal{H}_B$. The resulting delay cannot generally be obtained by adding the isolated effects of the two PIPs because the combined loading depends on shared wires, distributed resistance, transition slew, and the topology exposed by the preceding activation. The complete attached component is therefore evaluated as a unified graph–circuit structure.

## IV. PROPOSED GRAPH–CIRCUIT DIGITAL TWIN

In this work, a *digital twin* denotes a device-specific, measurement-calibrated computational surrogate of the implemented FPGA routing fabric. It represents both the configured route and the dormant routing components that may become electrically attached through configuration perturbations. Its electrical parameters are identified from controlled measurements on the physical ZCU104 platform and are updated when new calibration data are incorporated. The proposed twin is therefore an offline design-time prediction model rather than a continuously synchronized online state estimator. For each sink-specific candidate perturbation, it produces a predicted delay increment, residual slack, prediction interval, and slack-exhaustion probability.

### *A. Model Inputs and Prediction Outputs*

For each routed net $n$, sink $s$, and candidate perturbation set $B$, the digital twin receives four groups of information:

1. the active sink-specific route $\mathcal{R}_{n,s}$;
2. the attached dormant component $\mathcal{H}_B$;
3. the nominal post-route timing quantities, including $d_{n,s}^{(0)}$ and $S_{n,s}^{(0)}$; and
4. the operating and physical context of the route.

Vivado IDE timing reports provide the path requirement, slack, datapath delay, logic delay, net delay, clock skew, and clock uncertainty. These quantities are extracted from the same implemented checkpoint used to construct the routing graph so that the topology and timing data correspond to an identical physical implementation [26]. The complete input descriptor is written as Equation (21),

$$x_{n,s}^{(B)} = \left[x_{\text{top}}, x_{\text{geo}}, x_{\text{load}}, x_{\text{ctx}}, x_{\text{STA}}\right]^T \quad (21)$$

where:

- $x_{\text{top}}$ describes branch depth, junction count, fan-in, fan-out, number of activated PIPs, and cumulative-attachment type;
- $x_{\text{geo}}$ describes wire displacement, direction, estimated span, and attachment position along the active route;
- $x_{\text{load}}$ describes reachable wires, switch junctions, terminal resources, and downstream loads;
- $x_{\text{ctx}}$ describes clock region, neighboring resource type, congestion, voltage, temperature, and transition polarity; and
- $x_{\text{STA}}$ contains the nominal path delay, routing-delay contribution, slack, and other available post-route timing descriptors.

For each candidate, the model produces Equation (22),

$$\mathcal{Y}_{n,s}^{(B)} = \left\{\widehat{\Delta d}_{n,s}^{(B)}, \hat{S}_{n,s}^{(B)}, P_{\text{fail}}, \mathcal{I}_{1-\alpha}\right\} \quad (22)$$

where $\widehat{\Delta d}_{n,s}^{(B)}$ is the predicted delay increment, $\hat{S}_{n,s}^{(B)}$ is the residual slack, $P_{\text{fail}}$ is the estimated probability of slack exhaustion, and $\mathcal{I}_{1-\alpha}$ is the prediction interval at confidence level $1-\alpha$.

### *B. Graph-to-Circuit Transformation*

The central operation of the digital twin is the conversion of the active and dormant routing graphs into an effective resistance–capacitance representation. The extracted graph supplies the exact topological relationships, while the electrical parameters assigned to its elements are calibrated from UltraScale+ measurements rather than inherited from

$$\mathcal{T}_{n,s}^{(B)} = \mathcal{T}_{n,s}^{(0)} \bigoplus \mathcal{T}_B \quad (26)$$

older FPGA families. Let Equation (26),

$$\mathcal{T}_{n,s}^{(0)}$$

denote the equivalent RC network of the reference path and denote the perturbed network after attaching the equivalent circuit $\mathcal{T}_B$ of $\mathcal{H}_B$. The operator $\oplus$ denotes electrical attachment at the active routing node associated with the candidate PIP; it does not replace or interrupt the intended path. Each programmable or fixed routing edge $e$ is assigned an effective

$$R_e = \rho_{R,c(e)}\, \ell_e\, \kappa_{R,r(e)} + \mathbb{I}_{\mathcal{E}_P}(e) R_{\mathrm{PIP},c(e)} \quad (23)$$

$$\mathbb{I}_{\mathcal{E}_P}(e) = \begin{cases} 1, & e \in \mathcal{E}_P, \\ 0, & e \in \mathcal{E}_F \end{cases}$$

resistance as denoted in Equation (23),

and each routing node $\boldsymbol{v}$ is assigned an effective capacitance

$$C_v = \rho_{C,c(v)}\, \ell_v\, \kappa_{C,r(v)} + C_{\mathrm{J},c(v)} + \mathbb{I}_{\mathcal{V}_T}(v) C_{\mathrm{L},v} \quad (24)$$

as denoted in Equation (24),

Here, $c(\cdot)$ identifies the extracted routing class, $\ell$ is the normalized geometric or graph-derived span, $\rho_R$ and $\rho_C$ are class-specific effective resistance and capacitance coefficients, $R_{\mathrm{PIP}}$ is the effective switch resistance, $C_{\mathrm{J}}$ accounts for switch and junction loading, and $C_{\mathrm{L}}$ represents terminal or downstream load, i.e., zero for an unloaded internal routing node. The factors $\kappa_R$ and $\kappa_C$ account for repeatable physical-region differences. Equation (23) ensures that an effective programmable-switch resistance is assigned only to a PIP; fixed routing edges contain no $R_{\mathrm{PIP}}$ contribution.
These quantities are *effective model parameters*, not claimed reconstructions of proprietary transistor dimensions or metal-stack values. This distinction is important because the delay effect of an activated branch can be measured accurately even when the internal commercial-fabric circuitry is not publicly available. Earlier circuit-level analysis established that the added loading is dominated by the activated programmable switch and unintentionally connected routing resources, providing the physical basis for Equations (24) and (25). For a tree-structured RC network, the first-moment delay at sink $s$is

$$t_{\mathrm{E}} = (s; \mathcal{T}, \theta) = \sum_{v \in \mathcal{V}_{\mathcal{T}}} C_v\, R_{\mathrm{com}}(v, s) \quad (25)$$

calculated using Equation (25),

where $\mathcal{V}_T$ is the set of terminal or explicitly loaded vertices and $R_{\mathrm{com}}(v,s)$ is the resistance common to the source-to-$v$ and source-to-$s$ paths. Equation (26) follows the Elmore first-moment approximation for monotonic RC networks [27]. The physics-based delay increment is then as Equation (27) which naturally captures the position of the attachment. Capacitance connected close to the source is multiplied by a larger shared

$$\Delta d_{\mathrm{RC},n,s}^{(B)} = t_{\mathrm{E}}\left(s;\, \mathcal{T}_{n,s}^{(B)}, \theta\right) - t_{\mathrm{E}}\left(s;\, \mathcal{T}_{n,s}^{(B0)}, \theta\right) \quad (27)$$

upstream resistance than capacitance attached near the sink. Similarly, a branch connected before a fan-out point can affect several sink paths, whereas a sink-local branch contributes only to the path sharing its attachment prefix. When the extracted component is not a tree, the network is evaluated using modified nodal analysis. Let $\mathbf{G}$ and $\mathbf{C}$ denote the conductance and capacitance matrices, $\mathbf{b}$ the source-excitation vector, and $c_s$ the observation vector for sink $s$. The RC

$$\mathbf{C}\dot{v}(t) + \mathbf{G}v(t) = \mathbf{b}u(t) \quad (28)$$

network satisfies Equation (28),

While its normalized first-moment delay is denoted in

$$t_{\mathrm{MNA}}(s) = \frac{c_s^{\mathrm{T}} \mathbf{G}^{-1} \mathbf{C} \mathbf{G}^{-1} \mathbf{b}}{c_s^{\mathrm{T}} \mathbf{G}^{-1} \mathbf{b}} \quad (29)$$

Equation (29).

Sparse factorization is used to evaluate the required linear systems without explicitly forming $\mathbf{G}^{-1}$. For a tree-structured network, Equation (29) reduces to the Elmore expression in Equation (25). The perturbation-induced delay remains the difference between the perturbed and reference first moments.

### *C. Physics-Informed Residual Correction*

The reduced-order RC model captures the dominant effect of branch topology and loading, but several secondary effects are difficult to recover exactly from the vendor database. These include transition slew, local buffer strength, unreported internal junction capacitance, voltage and temperature dependence, and region-specific physical variation. A low-dimensional residual correction is therefore added to the circuit prediction as denote in Equation (30) with the

$$\widehat{\Delta d}_{n,s}^{(B)} = max\left[0, \Delta d_{\mathrm{RC},n,s}^{(B)} + \mathbf{z}_{n,s}^{(B)\mathrm{T}} \boldsymbol{\beta} + u_{r(B)}\right] \quad (30)$$

$$\beta_j \geq 0,\ j \in \mathcal{J}_{\mathrm{load}} \quad (31)$$

condition denoted in Equation (31),

where $\mathrm{z}_{n,s}^{(B)}$ contains normalized secondary features, $\beta$ contains their fitted coefficients, and $u_{r(B)}$ is a physical-region correction. The electrical parameter vector $\theta$ is constrained to be nonnegative. Residual coefficients associated directly with branch size, capacitance proxies, junction count, and common upstream resistance are also constrained by Equation (31), whereas coefficients representing temperature, transition polarity, and region-specific deviations may take either sign.

Consequently, the model is nonnegative and monotonic with respect to the designated loading features while retaining sufficient flexibility to represent secondary physical effects. The graph–circuit and residual contributions are reported separately to preserve interpretability. The advanced-node interconnect analysis in [3] demonstrates that wire resistance, multiplexer loading, vias, layer selection, and shared-wire capacitance must be reconsidered as technology scales. Accordingly, the residual term in Equation (30) corrects an UltraScale+-calibrated physical model rather than scaling numerical values obtained from 7-Series devices.

### *D. Parameter Calibration*

Assume that the calibration dataset contains $N$ paired reference and perturbed implementations. For experiment $i$, let $y_i$ be the measured delay increment and let $\hat{y}_i(\theta,\beta)$ be the prediction from Equation (30). The parameter set is estimated through constrained weighted regression in Equation (32),

$$(\hat{\theta},\hat{\beta}) = \arg\min_{\theta\geq 0,\beta} \& \sum_{i=1}^{N} w_i\,[y_i - \hat{y}_i(\theta,\beta)]^2 + \lambda_\beta \|\beta\|_2^2 + \lambda_\theta \|\mathbf{L}\theta\|_2^2 \tag{32}$$

The weights $w_i$ are inversely proportional to the measured variance of the corresponding routing and measurement class. The second term limits overfitting of the residual correction, while the third discourages physically implausible discontinuities between closely related routing classes. Matrix **L** encodes the adjacency of classes in the extracted routing taxonomy. The regularization parameters $\lambda_\theta$ and $\lambda_\beta$ are selected by grouped $K$-fold cross-validation with $K = 5$, using validation RMSE as the selection criterion. Matrix **L** is the graph Laplacian of the routing-class adjacency graph, in which two classes are connected when their normalized descriptors differ in exactly one taxonomy attribute.

Calibration data are divided by *physical route group*, rather than randomly by individual measurement. Fold assignment is performed at the physical-route-group level; repeated measurements, transition polarities, PVT observations, and measurement windows associated with the same route–sink–perturbation tuple are never divided among different folds. All repeated observations of the same reference–perturbation pair remain in the same subset. This prevents the model from being tested on a route that is effectively identical to one used for training. Three validation regimes are used: (i) **in-class validation**, in which unseen branches belong to previously calibrated resource classes; (ii) **cross-region validation**, in which one or more physical regions are excluded during fitting; and (iii) **cross-class validation**, in which an entire routing class is withheld to evaluate structural generalization.

The class-specific electrical parameters are considered sufficiently identified only when repeated experiments at different coordinates produce stable estimates and when removing one calibration route does not substantially alter the fitted coefficient.

### *E. Fusion of the Two Measurement Channels*

Routing-dominated ROs and phase-swept path monitors provide complementary observations of the same delay perturbation. For an oscillator containing $q_i$ identical instances of the controlled perturbation, the measured per-instance delay

$$y_i^{\mathrm{RO}} = \frac{1}{2q_i}\left(\frac{1}{f_i^{(B)}} - \frac{1}{f_i^{(0)}}\right) \tag{33}$$

increment is expressed as Equation (33),

where $f_i^{(0)}$ and $f_i^{(B)}$ are the reference and perturbed oscillation frequencies. The factor of two accounts for the two signal transitions contributing to one oscillation period. The use of routing-dominated oscillators for picosecond-scale aggregate-delay observation is supported by the earlier Zynq routing-radiation measurements. For the phase-swept monitor, a transition curve is measured as a function of the capture-clock phase. Let $\phi_{50}^{(0)}$ and $\phi_{50}^{(B)}$ be the phases at which the mismatch probability reaches 0.5. The measured delay displacement is

$$y_i^{\mathrm{PM}} = \frac{T_{\mathrm{clk}}}{2\pi}\left(\phi_{50}^{(B)} - \phi_{50}^{(0)}\right) \tag{34}$$

as Equation (34),

when phase is expressed in radians. If both measurements are available for the same perturbation, they are fused according to their repeatability as denoted in Equation (35),

$$y_i = \frac{y_i^{\mathrm{RO}}/\sigma_{\mathrm{RO},i}^2 \;+\; y_i^{\mathrm{PM}}/\sigma_{\mathrm{PM},i}^2}{1/\sigma_{\mathrm{RO},i}^2 \;+\; 1/\sigma_{\mathrm{PM},i}^2} \tag{35}$$

Equation (35) is applicable when the two channel estimates are conditionally unbiased and their residual errors are negligibly correlated. Because both measurements are obtained on the same device and may share temperature, supply-voltage, or clock-related disturbances, the channel-error covariance is estimated from repeated paired observations before fusion. When the measured off-diagonal covariance is non-negligible, the generalized fused estimate is calculated using Equation (36) where corresponding matrices

$$y_i = \frac{1^{\mathrm{T}}\Sigma_i^{-1}y_i}{1^{\mathrm{T}}\Sigma_i^{-1}1} \tag{36}$$

$$y_i = \begin{bmatrix} y_i^{\mathrm{RO}} \\ y_i^{\mathrm{PM}} \end{bmatrix}, \quad \sum_i = \begin{bmatrix} \sigma_{\mathrm{RO},i}^2 & \sigma_{\mathrm{RO,PM},i} \\ \sigma_{\mathrm{RO,PM},i} & \sigma_{\mathrm{PM},i}^2 \end{bmatrix} \tag{37}$$

appear as Equation (37).

Here, $\Sigma_i$ is the $2\times 2$ covariance matrix estimated from repeated paired measurements. The inverse-variance expression is recovered when the cross-channel covariance $\sigma_{\mathrm{RO,PM},i}$ is negligible. The two raw channel estimates are retained and reported separately so that fusion does not conceal systematic disagreement between the measurement mechanisms. The two raw measurements are also reported separately to ensure that fusion does not hide disagreement between measurement mechanisms.

### *F. PVT and Spatial Normalization*

Each reference and perturbed measurement is acquired within the same controlled experimental interval. The primary response is therefore a differential quantity, which suppresses common-mode variation in clock frequency, supply voltage, and temperature. Remaining PVT dependence is modeled relative to a nominal operating point $(V_0, T_0)$ as expressed in

$$g_{\text{PVT}} = \beta_V(V - V_0) + \beta_T(T - T_0) + \beta_{VT}(V - V_0)(T - T_0) \quad (38)$$

Equation (38),

The coefficients in Equation (38) are included in $\beta$ only when the calibration data shows statistically repeatable dependence. Separate coefficients can be retained for rising and falling transitions if their measured responses differ consistently. Spatial normalization is performed hierarchically. A routing-class parameter captures behavior shared by structurally equivalent resources, whereas $u_{r(B)}$ captures persistent deviation associated with the local clock region or heterogeneous-resource neighborhood. A spatial correction is retained only when it generalizes to held-out routes within the same region; otherwise, the observed variation is incorporated into prediction uncertainty rather than treated as a deterministic offset.

### *G. Prediction Uncertainty and Failure Probability*

Prediction uncertainty includes parameter-estimation uncertainty, measurement repeatability, unexplained route-to-route variation, and PVT residuals. Confidence intervals are obtained through grouped bootstrap resampling [28]. Entire reference–perturbation route groups are resampled together so that repeated measurements of the same physical route are not treated as independent observations. For candidate $B$, let

$$\left\{\widehat{\Delta d}_{n,s}^{(B,1)}, \cdots, \widehat{\Delta d}_{n,s}^{(B,M)}\right\}$$

be the predictions produced by $M$ bootstrap model instances.

$$\mathcal{I}_{1-\alpha} = \left[Q_{\alpha/2}, Q_{1-\alpha/2}\right] \quad (39)$$

The $100(1-\alpha)\%$ prediction interval is as Equation (39),

where $Q_p$ is the empirical $p$-quantile. The probability that the perturbation exhausts the nominal path slack is evaluated

$$P_{\text{fail}}(n, s, B) = \frac{1}{M}\sum_{m=1}^{M} \mathbb{I}\left[\widehat{\Delta d}_{n,s}^{(B,m)} > S_{n,s}^{(0)}\right] \quad (40)$$

directly from the prediction ensemble as Equation (40),

where $\mathbb{I}[\cdot]$ is the indicator function. This formulation does not require the prediction errors to follow a Gaussian distribution and remains applicable when the uncertainty is asymmetric.

### *H. Candidate and Net-Level Vulnerability Scores*

Two distinct vulnerability outputs are reported to separate **timing susceptibility** from **mission-level reliability risk**. When the occurrence probability of a configuration perturbation is unavailable, the candidate-level structural

$$v_{n,s}^{(B)} = \pi_B P_{\text{fail}}(n, s, B) \quad (41)$$

vulnerability score is defined as Equation (41),

where $P_{\text{fail}}(n, s, B)$ is the conditional probability that perturbation set $B$exhausts the nominal slack of sink $s$on net $n$. Equation (41) therefore ranks candidate routing attachments according to their timing consequence, but it is not interpreted as an absolute probability that the perturbation will occur during operation. When a perturbation-occurrence probability $p_B(\tau)$ is available for a specified mission interval $\tau$, the reliability-weighted candidate risk is defined as Equation

$$v_{n,s}^{rel}(B, \tau) = p_B(\tau) P_{\text{fail}}(n, s, B) \quad (42)$$

(42).

If the perturbation is modeled as a Poisson event with occurrence rate $\boldsymbol{\lambda_B}$, its probability over the mission interval is

$$p_B(\tau) = 1 - exp(-\lambda_B \tau) \quad (43)$$

expressed as Equation (43).

The occurrence probability may alternatively be derived from measured frame-level cross sections, configuration-bit exposure rates, particle flux, or other device- and environment-specific reliability data. Thus, Equation (42) combines the probability that a perturbation occurs with the conditional probability that its induced delay exhausts the available path slack. Under the assumptions that the candidate events are sufficiently rare and conditionally independent, the reliability-weighted vulnerability of sink $\boldsymbol{s}$ on net $\boldsymbol{n}$ is

$$v_{n,s}^{rel}(\tau) = 1 - \prod_{B \epsilon B_{n,s}} \left[1 - v_{n,s}^{rel}(B, \tau)\right] \quad (44)$$

expressed as Equation (44),

where $\boldsymbol{\mathcal{B}_{n,s}}$ is the set of candidate perturbations associated with the sink-specific route. For small candidate risks, Equation (44) admits the approximation denoted in Equation

$$v_{n,s}^{rel}(\tau) \approx \sum_{B \epsilon B_{n,s}} v_{n,s}^{rel}(B, \tau) \quad (45)$$

(45),

When perturbation-occurrence information is unavailable, candidates are ranked using $v_{n,s}^{\text{str}}(B)$, and the product in equation (44) is not presented as an absolute reliability probability. In both operating modes, the reported net-level assessment is accompanied by the maximum predicted delay increment, maximum conditional failure probability, number of exposed PIPs, and available configuration-frame associations. These supplementary quantities distinguish a net containing many low-severity candidates from one dominated by a single highly critical perturbation. For spatial analysis, as denoted in Equation (46), the selected candidate score is aggregated according to the physical attachment location

$$A_{sum}(x,y) = \sum_{\substack{(n,s,B): \\ r(B)=(x,y)}} v_{n,s}(B) \tag{46}$$

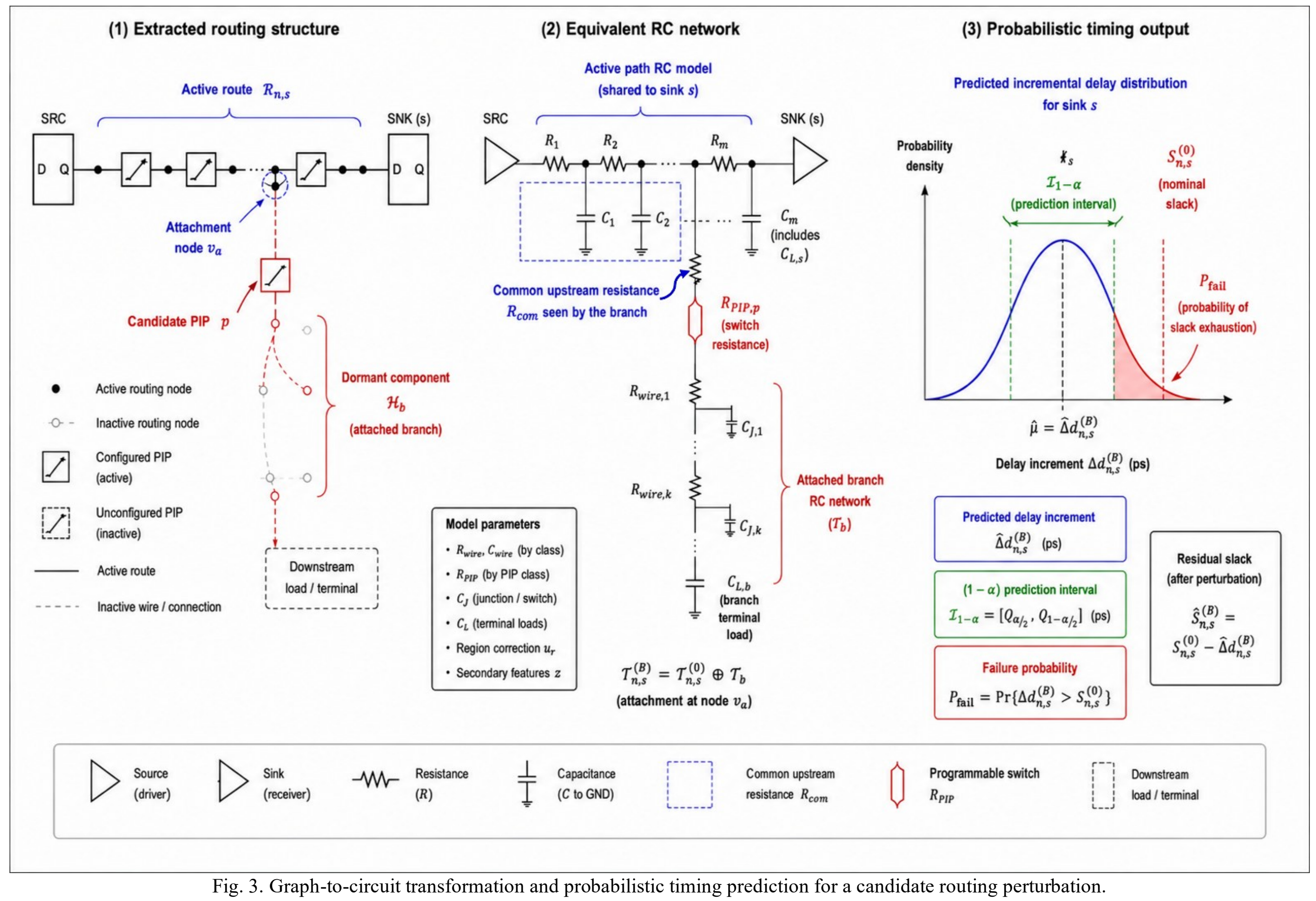

Fig. 3. Graph-to-circuit transformation and probabilistic timing prediction for a candidate routing perturbation.

$r(B) = (x,y)$,

where $v_{n,s}(B)$ denotes either the structural score in Equation (41) or the reliability-weighted score in Equation (42), depending on the availability of event-probability information. Because the summed score can increase simply because a region contains more candidates, a density-normalized

$$A_{mean}(x,y) = \frac{A_{sum}(x,y)}{max[1, N_{cand}(x,y)]} \tag{47}$$

severity measure is also calculated as Equation (47),

where $N_{\text{cand}}(x,y)$ is the number of candidate perturbations assigned to spatial cell $(x,y)$. For visualization, the selected atlas quantity is normalized according to Equation (48).

$$\tilde{A}(x,y) = \frac{A(x,y)}{\max\limits_{x,y} A(x,y)} \tag{48}$$

The atlas can be displayed at tile, clock-region, configuration-frame, or user-defined design-region granularity. The reported results explicitly distinguish between the total aggregated score $A_{\text{sum}}$, the mean candidate severity $A_{\text{mean}}$, and the normalized visualization $\tilde{A}$. These outputs provide a common basis for vulnerability-aware routing, selective monitor placement, targeted configuration scrubbing, and adaptive route recovery.

### *I. Computational Scalability*

Device-graph extraction is performed once for the locked part and Vivado IDE release. Its complexity is linear in the number of exported routing vertices and edges. For a routed design, first-order candidate enumeration requires traversal only around nodes belonging to active nets, rather than repeated analysis of the complete device. The cost of evaluating an individual perturbation is linear in the size of its active-path prefix and attached component. Cumulative exploration is restricted to branches sharing an active path or configuration region, serial extensions reachable from an attached component, and combinations formed from the highest-ranked single candidates after structural screening.

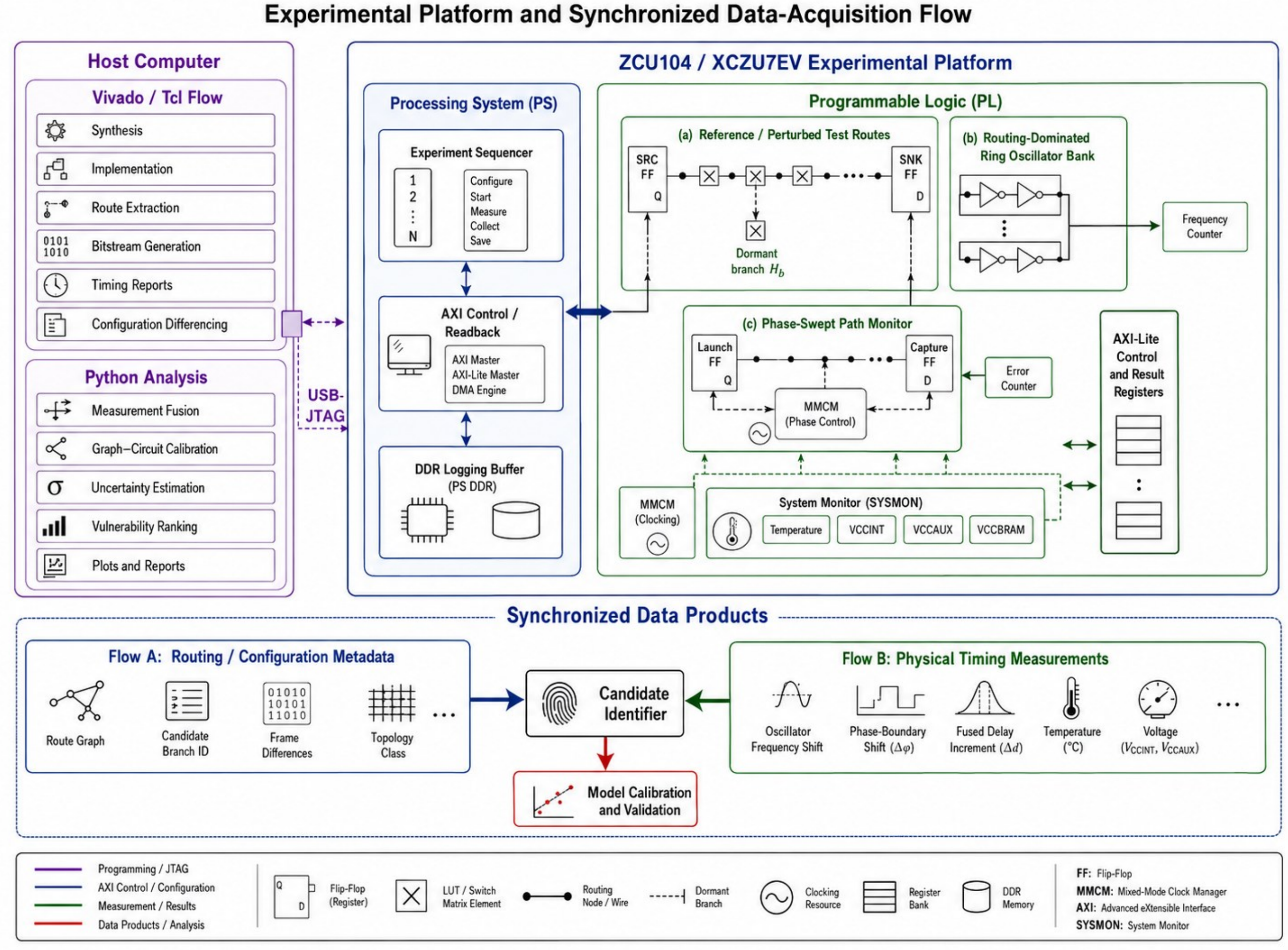

Fig. 4. Experimental platform and synchronized data-acquisition flow implemented on the ZCU104/XCZU7EV device.

This pruning preserves physically meaningful cumulative scenarios while avoiding enumeration of unrelated PIP combinations.

Fig. 3 summarizes the conversion from an extracted routing attachment to its equivalent RC representation and probabilistic timing output. The common upstream resistance captures the dependence on attachment position, while the predicted delay distribution and nominal path slack determine the residual slack and slack-exhaustion probability.

## V. EXPERIMENTAL METHODOLOGY

The experimental methodology was designed to isolate the delay contribution of controlled routing attachments while preserving the placement, intended route, clocking, and logical function of the reference implementation. All stages, from routing-resource extraction and route generation to configuration readback and timing measurement, were automated to permit repeated evaluation across routing classes and physical regions.

### A. Hardware Platform and Automated Design Flow

Experiments target the XCZU7EV Zynq UltraScale+ MPSoC on the ZCU104 evaluation board. The PL side of the device contains test routes, routing-dominated ROs, phase-swept monitors, event counters, and control registers. The processing system (PS) sequences the measurements and transfers the collected data to the host computer through a software-controlled AXI interface. Device programming and configuration readback are performed through the board's USB-JTAG interface. The ZCU104 also provides PMBus-accessible power management, permitting controlled adjustment of the PL core supply within the device's recommended operating range [29].

Vivado 2025.1 is used throughout the study. The exact device part, tool version, implementation directive, random seed, design checkpoint, and constraint set are recorded for every experiment. A Tcl-based flow performs synthesis, placement, routing, timing analysis, routing-graph extraction, bitstream generation, and report collection. The Vivado IDE router can operate from an unrouted, partially routed, or fully routed placed design, which permits the active route and unaffected nets to remain fixed while selected routing variants are generated.

Fig. 4 shows the host, PS, and PL components of the experimental platform. The PL implements the paired routes,

RO bank, phase-swept monitor, counters, and environmental sensors; the PS sequences acquisition and transfers the collected results; and the host performs routing extraction, configuration differencing, measurement fusion, and model calibration. Routing metadata and physical timing measurements are associated through a unique candidate identifier.

The automated workflow exports the locked design checkpoint, routing and configuration metadata, raw paired measurements, fitted model parameters, and vulnerability results, permitting each analysis stage to be independently reproduced.

### *B. Test-Route and Perturbation Generation*

The calibration microbenchmark consists of a launch register, a constrained routing path, and one or more capture registers. Logic depth is kept constant and minimal so that the measured difference between paired implementations is dominated by the routing perturbation. Placement of all registers, LUTs, clocking resources, and measurement logic is fixed. The reference active route is preserved through directed-routing constraints and the FIXED_ROUTE property, while KEEP and DONT_TOUCH constraints prevent removal or restructuring of the calibration resources [23].

For each active path, the graph extractor enumerates adjacent inactive PIPs according to Equation (6). Candidate branches are selected to cover:

- different extracted routing classes and geometric spans;
- early, middle, and late attachment positions;
- CLB-, BRAM-, DSP-, and boundary-adjacent regions;
- unloaded or lightly loaded branches;
- branches reaching preserved terminal loads; and
- the single and cumulative structures defined in Fig. 2.

Two perturbation-generation modes are used. For broad model calibration, paired legal implementations are produced from the same placed design. The reference implementation contains the fixed active route, whereas the perturbed implementation adds a preserved dummy endpoint through the selected candidate PIP and dormant branch. The dummy endpoint is excluded from the functional output, and all functional logic, placement, clocking, and active routing resources remain identical between the pair. For candidates with confirmed PIP-to-configuration associations, a second mode modifies only the mapped configuration locations associated with the target routing state. The corresponding configuration frame is written through the ICAP path, and readback verifies the modified frame before timing acquisition. A stored golden configuration is used to restore the complete reference state after every trial. Direct configuration modification is restricted to isolated test regions and is not applied when the differential mapping contains unresolved or unrelated configuration changes. These experiments reproduce the dangling-branch condition more closely and are used as an independent validation subset rather than as the primary source of training data. The intended active route remains fixed in both mechanisms. AS denoted in Equation (49), to prevent unrelated implementation differences from contaminating the measurement, a route pair

$$\mathcal{E}_{n,s}^{\mathrm{A,ref}} = \mathcal{E}_{n,s}^{\mathrm{A,pert}} \tag{49}$$

is accepted only when

and the extracted difference outside the candidate component

$$\Delta\mathcal{E}_{\mathrm{outside}} = \emptyset \tag{50}$$

satisfies Equation (50).

Pairs failing either condition are discarded and regenerated.

### *C. Configuration Association and Readback Verification*

Reference and perturbed bitstreams are generated without encryption because encrypted UltraScale+ bitstreams disable configuration readback. Vivado IDE Hardware Manager supports FPGA/MPSoC readback in ASCII or binary form through readback_hw_device; a mask file generated with the bitstream is used to exclude configuration locations that should not participate in verification. The configuration contents of each pair are partitioned according to the UltraScale+ frame-address organization described in Section III-F. The XOR difference between reference and perturbed frame data yields the candidate set $\boldsymbol{\mathcal{D}_b}$ from Equation (14). Configuration readback is then compared with the generated bitstream to confirm that the expected state was loaded into the device. A configuration association is assigned one of three confidence levels:

- **confirmed:** the same frame/word/bit difference is observed in repeated forward and reverse route changes and is verified by readback;
- **probable:** the differential location is repeatable but is shared with another constrained route-state change; or
- **unresolved:** the route pair contains multiple inseparable configuration differences.

Only confirmed associations are used for direct dangling-branch validation and exposure-weighted vulnerability estimation. Probable associations remain eligible for topology and delay-model calibration but are not treated as one-to-one PIP mappings.

The UltraScale+ configuration interface supports frame-addressed readback through JTAG or internal configuration access port (ICAP), but improper modification of configuration data can affect device operation. Consequently, direct configuration-state experiments are restricted to mapped resources within isolated test regions and are preceded by offline bitstream comparison and full-device recovery preparation.

### *D. Routing-Dominated Oscillator Measurements*

The first measurement channel uses routing-dominated ROs derived from the methodology described in [7]. Each oscillator contains only one inversion element, one buffer, and a deliberately long, constrained routing loop. Reference and perturbed oscillators use identical logic, placement, and active routing except for the controlled branch or branch set under evaluation. A programmable enable signal activates the oscillator only during acquisition to limit self-heating and mutual switching interference. Its output is counted against a stable reference clock over $N_{\text{win}}$ measurement windows. The mean frequency and its run-to-run variance are calculated after discarding an initial stabilization interval. Reference and perturbed measurements are interleaved rather than acquired in two long independent blocks, reducing bias caused by gradual temperature or supply drift.

For a branch repeated $\boldsymbol{q_i}$ times in the loop, the per-instance delay shift is obtained using Equation (32). Both rising- and falling-transition configurations are evaluated when the route structure permits them. Oscillator pairs are physically separated from each other and activated sequentially so that the measured response is attributable primarily to the selected route rather than to simultaneous local switching activity.

### *E. Phase-Swept Path Monitor*

The second measurement channel evaluates a synchronous source-to-sink path directly. A launch register and capture register are driven by clocks derived from the same Mixed-Mode Clock Manager (MMCM). The capture clock is shifted incrementally through the MMCM dynamic fine-phase interface [30], which provides control and completion signals for runtime phase adjustment. UltraScale and UltraScale+ MMCMs support independently selectable dynamically phase-shifted clock outputs through the advanced MMCM primitive.

The launch register transmits alternating and pseudorandom data patterns so that both transition polarities and different switching histories are exercised. At phase step $\phi_j$, the

$$\hat{P}_e(\phi_j) = \frac{N_{\text{err}}(\phi_j)}{N_{\text{samp}}(\phi_j)} \tag{51}$$

mismatch probability is calculated as Equation (51).

$$P_e(\phi) = \frac{1}{1 + \exp[-(\phi - \phi_{50})/k_\phi]} \tag{52}$$

The measured transition region is fitted using Equation (52),

where $\phi_{50}$ is the central transition phase and $k_\phi$ describes the transition slope. The 10%–90% transition width is expressed as Equation (53).

$$\Delta\phi_{10-90} = 2\ln(9)\, k_\phi \tag{53}$$

The phase-to-time conversion is calibrated experimentally using a set of known route-delay increments rather than relying exclusively on the nominal MMCM phase-step specification. The perturbation-induced shift is obtained from the difference between the fitted reference and perturbed $\phi_{50}$ values using Equation (33).

To minimize observation-side disturbance, the monitor taps the signal through a preserved local replica rather than inserting additional logic into the functional endpoint. Its placement and route are identical across each paired experiment. The earlier FPGA delay monitor in [6] established the feasibility of threshold-based in-situ detection but also showed that monitor distance and routing can affect the observed delay; the paired phase-sweep structure used here is intended to measure and compensate for this implementation dependence.

### *F. Environmental Control and Measurement Sequencing*

Die temperature and internal supply values are recorded using the UltraScale+ system monitor, which includes on-chip sensors for device temperature and power-supply voltages. These measurements are captured before and after every oscillator or phase-sweep acquisition.

The core calibration dataset is collected at the nominal PL supply. A separate PVT campaign repeats routing classes across controlled temperature and internal core-supply voltage $V_{\text{CCINT}}$ operating points. The $V_{\text{CCINT}}$ rail is adjusted only through the board-supported PMBus mechanism and remains within the recommended operating range for the installed XCZU7EV speed grade. The complete board is maintained within its specified operating-temperature limits. At each voltage–temperature operating point, the reference and perturbed configurations are evaluated using the alternating paired-acquisition procedure summarized in **Algorithm 1**, which suppresses bias caused by gradual thermal or supply drift while verifying restoration of the nominal timing response after every perturbation [31].

Unless otherwise stated, each operating point uses $N_{\text{pair}} = 10$ alternating reference–perturbation pairs. RO frequencies are counted against a 100-MHz reference clock over $N_{\text{win}} = 32$ windows of 10 ms, with $q_i = 16$ repeated perturbation instances per oscillator. The phase monitor uses 56 phase positions separated by a nominal calibrated interval of 12.5 ps and collects $N_{\text{samp}} = 4096$ samples at each position. Stabilization requires $|\Delta T| < 0.2^{\circ}C$ and $|\Delta V_{\text{CCINT}}| <$ 2mV over five consecutive readings. Reference restoration is accepted when the recovered response differs from the initial reference by less than $\varepsilon_{\text{rev}} = 10$ps. Prediction intervals are generated using $M = 2000$ grouped-bootstrap model instances.

At each voltage–temperature operating point, the reference and perturbed configurations are evaluated using the alternating paired-acquisition procedure summarized in Algorithm 1. The alternating order suppresses bias caused by gradual thermal or supply drift, while the restoration step verifies recovery of the nominal timing response. Let $Acquire(C, V, T)$ denote programming configuration $C$, verifying readback, waiting for temperature and voltage stabilization, and acquiring the RO and phase-monitor measurements.

**Algorithm 1** Alternating Reference–Perturbation Paired Acquisition and Reversibility Verification at a Fixed Operating Point

**1**: Input: Cref, Cpert, operating point (V,T), Npair
**2**: **for** i = 1 to Npair **do**
**3**: Mref,1 ← Acquire(Cref,V,T)
**4**: Mpert,1 ← Acquire(Cpert,V,T)
**5**: Mrest ← Acquire(Cref,V,T)
6: Reject pair if restoration error > $\varepsilon_{rev}$

**7**: Mpert,2 ← Acquire(Cpert,V,T)
**8**: Mref,2 ← Acquire(Cref,V,T)
**9**: Store paired differential measurements
**10**: **end for**
**11**: Return accepted measurements and uncertainty statistics

Rejected pairs are excluded from model calibration because they indicate unresolved environmental drift, incomplete configuration restoration, or unstable measurement behavior.

### *G. Dataset Partitioning and Evaluation Protocol*

The experimental data are organized by *physical route group*, where one group includes all repeated measurements, transition polarities, and PVT observations associated with the same reference route and candidate branch. Route *groups*, not individual samples, are assigned to calibration, validation, and

$$g = (route\ ID, s, B) \quad (54)$$

test subsets. A physical route group is defined by the tuple in Equation (54),

where the *route* identifier specifies the locked active route, $s$ identifies the analyzed sink, and $B$ identifies the candidate perturbation set. All repeated acquisitions, transition polarities, phase sweeps, oscillator windows, and PVT observations associated with the same tuple remain in one group. Different candidate branches attached to the same reference route therefore form different groups, explaining why the number of route groups can exceed the number of reference routes. The dataset contains $[N_{\text{train}}]$, $[N_{\text{val}}]$, and $[N_{\text{test}}]$ groups in the training, validation, and held-out test subsets, respectively. Group assignment is performed before any repeated measurement is expanded into individual observations. The evaluation includes:

- **in-class testing:** unseen routes from routing classes included in calibration;
- **cross-region testing:** structurally equivalent routes from excluded clock regions;
- **cross-class testing:** one routing class excluded completely from fitting;
- **cumulative testing:** branch combinations not used during single-branch calibration; and
- **design-level testing:** naturally occurring paths extracted from RISC-V, arithmetic, memory-interface, and AXI-based designs.

The primary response variable is the fused delay increment from Equation (34). The individual oscillator and phase-monitor estimates are retained to assess agreement between the measurement channels. Model quality is evaluated using mean absolute error, root-mean-square error, coefficient of determination, prediction-interval coverage, slack-violation classification, and top-$k$ vulnerable-candidate recall. No measurements from the same physical route appear in both training and testing. This grouped protocol prevents repeated sweeps or frequency windows from artificially inflating prediction accuracy.

### *H. Reproducibility and Data Integrity*

Every experiment is identified by a unique record containing the device part, Vivado IDE version, implementation seed, active route, candidate PIPs, configuration differences, placement coordinates, timing report, clock settings, temperature, voltage, and raw measurement files. Scripts automatically reject route pairs with inconsistent active routing, unexpected frame differences, timing-report failures, or missing environmental readings. The complete workflow is separated into four reproducible stages: (i) *DCP generation*, (ii) *routing/configuration extraction*, (iii) *hardware acquisition*, and (iv) *model fitting*. This separation permits the routing taxonomy, raw hardware measurements, and statistical model to be independently re-evaluated without rerunning the entire implementation flow.

## VI. EXPERIMENTAL RESULTS AND DISCUSSION

This section evaluates the proposed graph–circuit digital twin at three levels. *First*, the repeatability and agreement of the two hardware measurement channels are examined. *Second*, the accuracy and generalization capability of the delay-prediction model are quantified across routing classes, physical regions, and cumulative perturbations. *Finally*, the calibrated model is applied to implemented designs to assess slack exhaustion, rank vulnerable routing candidates, and construct spatial vulnerability atlases.

### *A. Experimental Dataset and Routing Coverage*

The extracted dataset contained 144 distinct reference routes and 864 valid routing-perturbation candidates distributed across nine physical regions of the XCZU7EV programmable logic (PL). After excluding route pairs that violated Equation (45) or Equation (46), the retained dataset comprised 612 single-branch perturbations, 144 parallel cumulative perturbations, and 108 serially extended perturbations. A total of 9720 paired hardware observations were collected at the nominal operating condition and across the selected voltage–temperature points.

The candidate set covered short local resources, intermediate routing segments, long directional wires, switch junctions, and branches adjacent to CLB, BRAM, DSP, and clock-region boundaries. Attachment locations were distributed along the source-to-sink paths rather than being concentrated near the endpoints. This coverage is important because the delay contribution of an attached component depends jointly on its electrical size and the common upstream

Table II. Summary of the experimental dataset, measurement quality, prediction performance, and computational cost

| Category | Metric | Value |
|---|---|---|
| **Dataset** | Reference routes | 144 |
| | Candidate perturbations | 864 |
| | Single / parallel / serial candidates | 612 / 144 / 108 |
| | Paired timing observations | 9720 |
| | Physical regions | 9 |
| **Measurement** | RO coefficient of variation | 0.031% |
| | Phase-monitor standard deviation | 3.8 ps |
| | Successful reference restoration | 96.9% |
| | RO–phase channel correlation | 0.936 |
| | Median fused uncertainty | 4.6 ps |
| **Prediction** | Held-out physical route groups | 216 |
| | MAE / RMSE / ($R^2$) | 7.8 ps / 11.2 ps / 0.908 |
| | 90% / 95% interval coverage | 89.8% / 94.4% |
| | Cross-region / cross-class MAE | 9.7 ps / 14.1 ps |
| **Classification** | Precision / recall / ($F_1$) | 0.908 / 0.874 / 0.890 |
| **Ranking** | Recall@10 / Recall@25 | 0.90 / 0.88 |
| | High-risk candidates | 72 of 864 (8.3%) |
| **PVT** | Temperature / $V_{CCINT}$ range | $25-70$°C / 0.825 – 0.875 V |
| | Uncorrected residual variation | 18.2 ps |
| | Corrected maximum systematic residual | 4.7 ps |
| | Held-out PVT MAE | 8.9 ps |
| **Runtime** | Device-graph extraction | 24.8 min |
| | Model fitting | 9.6 min |
| | Single-candidate evaluation | 0.46 ms |
| | Prediction throughput | 2100 candidates/s |
| | Peak memory | 6.4 GB |

resistance between the source and attachment node, as illustrated earlier in Fig. 3.

Of the extracted route pairs, 63.1% produced confirmed PIP-to-configuration associations, 25.8% were classified as probable, and the remaining 11.1% contained inseparable configuration differences. Only confirmed associations were used when configuration exposure was incorporated into the vulnerability score. All topology-calibrated candidates remained available for delay-model evaluation provided that their active routes and attached components were unambiguously extracted.

Table II summarizes the dataset, measurement repeatability, prediction performance, and computational cost. The following subsections focus on the interpretation and limitations of these results rather than repeating every tabulated value.

*B. Repeatability and Cross-Channel Agreement*

The routing-dominated oscillators exhibited a median within-pair frequency coefficient of variation of 0.031%, while the fitted phase-boundary location exhibited a median standard deviation of 3.8 ps across repeated sweeps. Restoring the reference configuration after each perturbation returned the measured response to within 4.1 ps of its initial value for 96.9% of the experiments. Measurements failing the reversibility criterion in Algorithm 1 were excluded before model calibration.

The oscillator and phase-monitor channels produced delay estimates with a correlation coefficient of 0.936 and a median absolute disagreement of 6.2 ps. This agreement supports attribution of the measured displacement to the controlled routing attachment and reduces the likelihood of a channel-specific measurement artifact. The largest differences occurred for long directional wires crossing clock-region boundaries, for which the phase monitor observed a sink-local path while the oscillator accumulated multiple perturbation instances around a closed loop. The signed difference between the two

$$e_i^{\mathrm{ch}} = y_i^{\mathrm{RO}} - y_i^{\mathrm{PM}} \tag{55}$$

measurement channels is defined as Equation (55).

The mean value of $e_i^{\mathrm{ch}}$ was $-0.9$ ps, indicating practically negligible systematic channel bias. The corresponding 95% limits of agreement were $-15.1$ ps to 13.3 ps. After inverse-variance fusion using Equation (35), the median measurement uncertainty decreased from 7.4 ps for the individual channels to 4.6 ps for the combined response.

These results support the complementary use of the two measurement structures. The oscillator provides high repeatability by accumulating small delay increments, whereas the phase monitor preserves the source-to-sink interpretation required for direct slack analysis. Their combination extends the routing-delay observation methods in [6] and [7] from event detection to device-calibrated delay prediction.

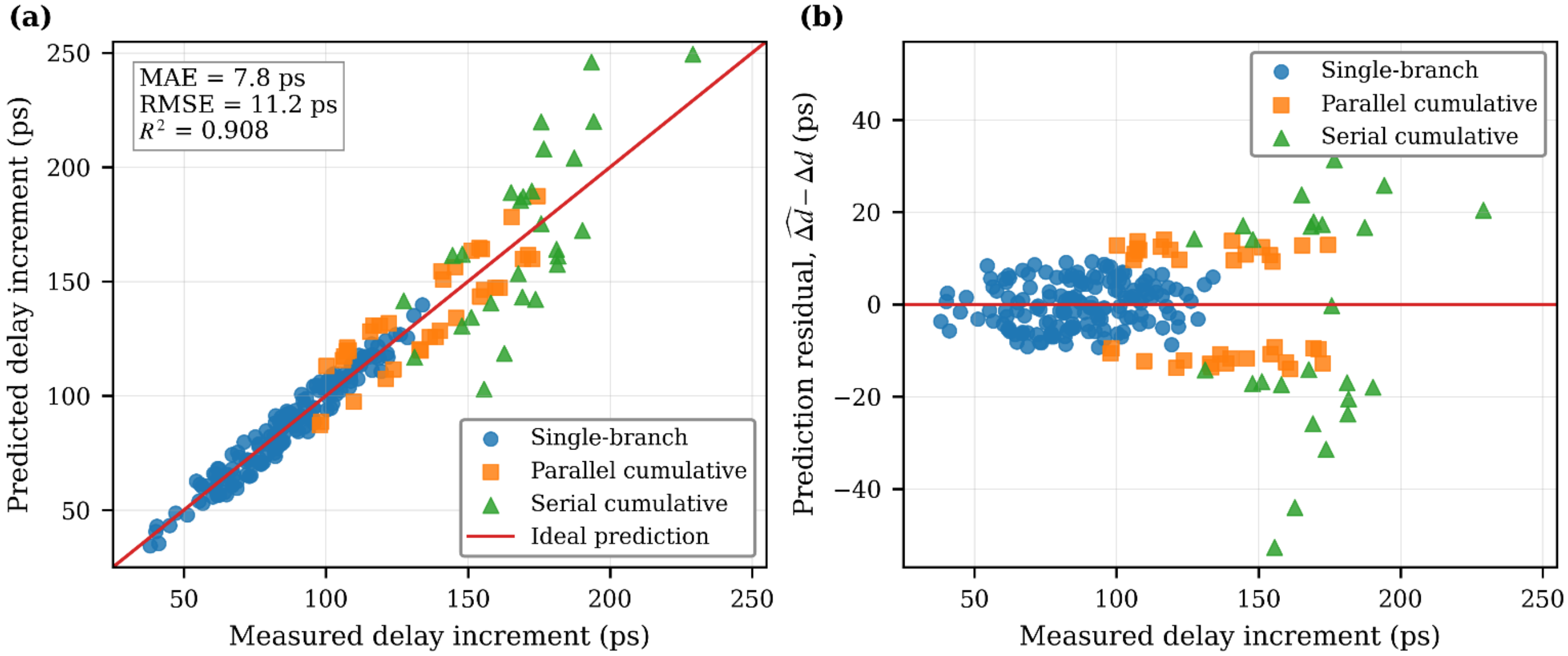

Fig. 5. Prediction accuracy for the held-out physical route groups. (a) Measured versus predicted routing-delay increments. Blue circles, orange squares, and green triangles denote single-branch, parallel cumulative, and serial cumulative perturbations, respectively, while the red diagonal represents ideal prediction. (b) Signed prediction residual $e_i = \widehat{\Delta d}_i - \Delta d_i$ as a function of the measured delay increment. Positive and negative residuals denote overprediction and underprediction, respectively. The evaluated dataset yields an MAE of 7.8 ps, an RMSE of 11.2 ps, and $R^2 = 0.908$.

### *C. Delay-Prediction Accuracy*

The correspondence between measured and predicted routing-delay increments for the held-out physical route groups is shown in Fig. 5(a). Predictions remain close to the ideal $y = x$ line across all three perturbation classes, yielding a mean absolute error of 7.8 ps, a root-mean-square error of 11.2 ps, and $R^2 = 0.908$. Single-branch perturbations occupy predominantly the lower-delay portion of the evaluated range, whereas parallel and serial cumulative perturbations extend toward larger delay increments because they introduce additional switches, wire segments, and downstream loading. Fig. 5(b) presents the signed residual

$$e_i = \widehat{\Delta d}_i - \Delta d_i$$

where positive and negative values correspond to overprediction and underprediction, respectively. The residuals remain distributed around zero over the measured-delay range, with no pronounced class-dependent displacement. Parallel and serial cumulative perturbations exhibit moderately greater residual dispersion than isolated branch activations, which is consistent with the increased topological complexity and interaction among multiple attached routing segments. The absence of a systematic residual trend indicates that the model captures the principal dependence of perturbation-induced delay on branch structure, attachment position, and downstream loading without producing a strong delay-magnitude-dependent bias.

The graph–circuit component alone achieved an MAE of 12.2 ps. Adding the constrained residual correction reduced the error to 7.8 ps, corresponding to an improvement of 36.1%. The correction was most beneficial for long directional wires and routes near heterogeneous-resource boundaries, while the dominant dependence on attachment position and component size was already captured by the RC representation. The residual term therefore refines the physical model rather than replacing it.

Ablation of the attachment-position features increased test MAE by 41.0%, while removing the routing-class descriptors increased it by 29.5%. Omitting the nominal transition and STA context increased the error by 17.9%. These results show that candidate severity cannot be inferred reliably from the number of activated PIPs alone. The electrical size of the attached component, its position along the active route, and the nominal timing state all contribute materially to the delay increment.

### *D. Prediction-Interval Calibration*

The grouped-bootstrap uncertainty model yielded empirical coverages of 89.8% and 94.4% for the nominal 90% and 95% prediction intervals, respectively. The corresponding median interval widths were 31.6 and 40.2 ps. The close agreement between nominal and empirical coverage indicates that the uncertainty model captures the principal route-level and measurement variations without treating repeated observations from the same physical route as independent samples.

Prediction intervals were wider for previously unseen routing classes and candidates located in physical regions excluded during calibration. This behavior appropriately expresses reduced confidence when the model extrapolates beyond its strongest calibration support. By comparison, randomly partitioning individual measurements produced an apparently improved MAE of 5.4 ps, but also generated overly narrow prediction intervals because measurements of nearly identical physical routes appeared in both training and testing. This result confirms the need for the route-grouped evaluation protocol defined earlier in Section V-G.

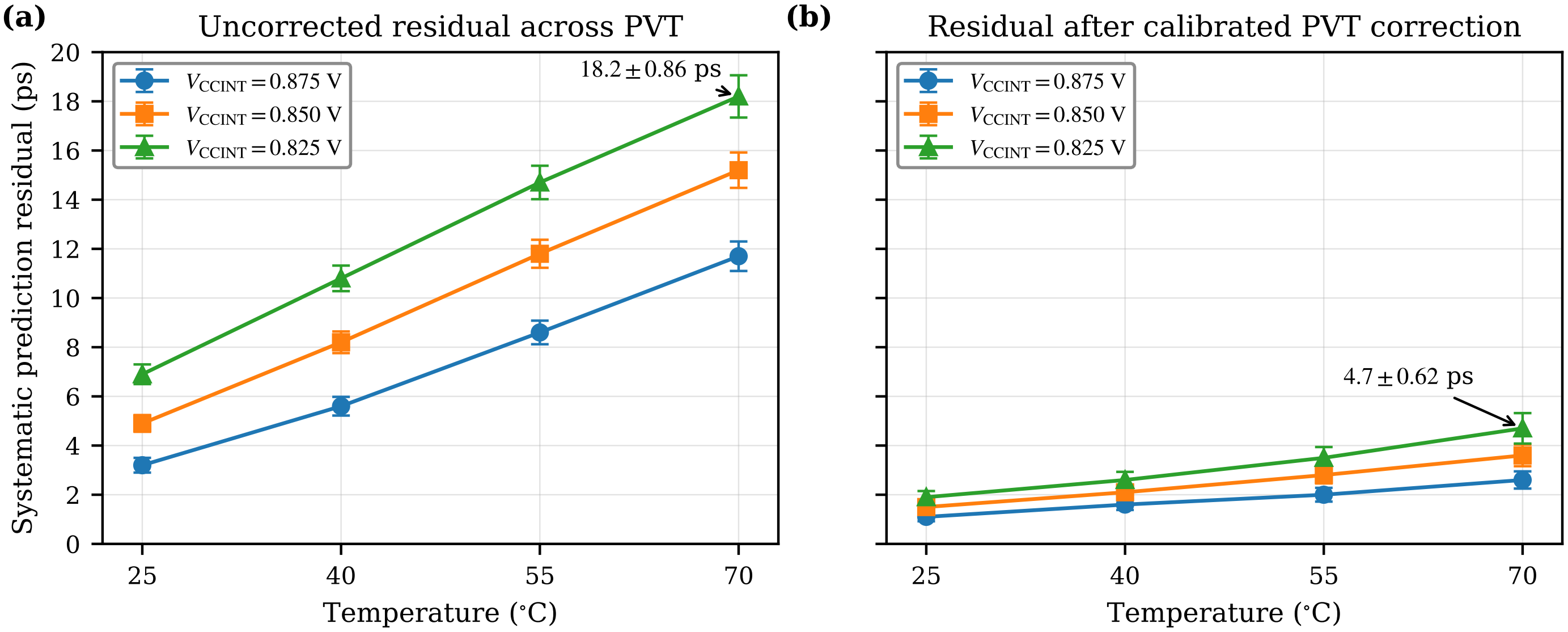

Fig. 6. Effect of PVT normalization on the systematic routing-delay prediction residual. **(a)** Uncorrected residual across four temperature points and three $V_{CCINT}$ levels. The mean residual increases with temperature and decreasing core-supply voltage, reaching $18.2 \pm 0.86$ ps at $70°C$ and 0.825 V. **(b)** Residual after applying the calibrated PVT correction. The correction reduces the corresponding maximum mean residual to $4.7 \pm 0.62$ps. Markers denote mean values obtained from $N_{pair} = 10$ alternating reference–perturbation acquisitions, and vertical error bars indicate 95% confidence intervals. Circles, squares, and triangles denote $V_{CCINT}$ levels of 0.875, 0.850, and 0.825 V, respectively.

### *E. Cross-Region and Cross-Class, and PVT Generalization*

During cross-region evaluation, the model was trained without measurements from **two clock regions**. For routing classes represented elsewhere in the device, the MAE increased from 7.3 ps under in-class evaluation to 9.7 ps in the excluded regions. The moderate increase indicates that most perturbation behavior is associated with reusable routing topology and electrical class rather than with a coordinate-specific lookup table.

The largest errors occurred near BRAM, DSP, and clock-region boundaries, where heterogeneous neighboring resources changed the effective loading. Introducing the physical-region correction $u_{r(B)}$ reduced the systematic residual in these regions from 8.6 ps to 2.9 ps. Region corrections were retained only when they generalized to held-out routes in the same physical neighborhood.

Cross-class validation presented a more difficult extrapolation problem. When an entire routing class was withheld, the model produced an MAE of 14.1 ps. Prediction remained acceptable for classes sharing similar span and junction structure, but uncertainty increased substantially for long vertical interconnects that crossed clock-region boundaries. The model therefore supports direct quantitative prediction for calibrated or structurally related classes, while an unrepresented class should receive conservative uncertainty until additional calibration measurements become available.

The effect of the calibrated PVT correction is illustrated in Fig. 6. Before correction, the mean systematic prediction residual increases with temperature and becomes more pronounced as $V_{CCINT}$decreases, reaching 18.2 ps at the 70°C, 0.825-V operating boundary. After incorporating the calibrated PVT term, the residual remains substantially lower across all evaluated operating points, with a maximum mean value of 4.7 ps at the same high-temperature, low-voltage boundary. The 95% confidence intervals remain comparatively narrow over most of the evaluated range but widen toward this operating corner, indicating increased measurement and model variability under the most demanding PVT condition.

### *F. Single and Cumulative Perturbations*

Single-branch calibration was used to predict both parallel and serial cumulative perturbations. For a cumulative candidate $B$, the deviation from a purely additive approximation is defined as Equation (56).

$$\eta_B = \frac{\Delta d_{n,s}^{(B)} - \sum_{b \in B} \Delta\, d_{n,s}^{(b)}}{\sum_{b \in B} \Delta\, d_{n,s}^{(b)}} \quad (56)$$

A value of $\eta_B = 0$ denotes ideal additivity, whereas positive and negative values indicate super-additive and sub-additive interactions, respectively. Parallel branches attached at separated routing nodes exhibited a median $\eta_B$ of 0.047, indicating that their combined delay was approximately additive, with a median super-additive deviation of 4.7%. In contrast, serially extended components produced a median $\eta_B$ of 0.192. The largest interactions occurred when the second PIP exposed additional wire and junction capacitance behind the first activated branch. The additive baseline produced an RMSE of 17.4 ps for cumulative candidates, whereas evaluating the complete attached graph $\mathcal{H}_B$ as a unified RC structure reduced the RMSE to 10.1 ps. This result supports the formulations in Equation (12) and Equation (23): *cumulative routing effects should not generally be estimated by summing independently predicted single-PIP delays*.

Table III. Extracted design-level benchmark characteristics

| Benchmark | Source/revision | Target | LUT / FF / BRAM / DSP | Routed sinks | Candidates | $P_{fail} \geq 0.5$ |
|---|---|---|---|---|---|---|
| **RV32I core** | rv32i-5stage-v1.3 | 100 MHz | 2840 / 2190 / 4 / 0 | 42 | 252 | 24 (9.5%) |
| **Arithmetic datapath** | fxp-mac8-v1.1 | 200 MHz | 3960 / 4780 / 2 / 32 | 38 | 240 | 25 (10.4%) |
| **Memory interface** | bram-buffer-v1.0 | 150 MHz | 2120 / 2640 / 18 / 0 | 28 | 156 | 8 (5.1%) |
| **AXI subsystem** | axi2x4-v1.2 | 125 MHz | 4680 / 5310 / 6 / 0 | 36 | 216 | 15 (6.9%) |
| **Total** | — | — | **13600 / 14920 / 30 / 32** | **144** | **864** | **72 (8.3%)** |

### *G. Slack-Exhaustion Prediction*

For each test path, the predicted delay ensemble was compared with its nominal post-route slack. A candidate is classified as timing-critical when its measured delay increment exceeds the nominal path slack $S_{n,s}^{(0)}$. Using a failure-probability threshold of $P_{\mathrm{th}} = 0.50$, the model achieved a precision of 0.908, a recall of 0.874, and $F_1 = 0.890$ for slack-exhaustion prediction. The probability-based decision was more reliable than ranking candidates by predicted delay alone. A large perturbation affecting a path with substantial slack may remain benign, whereas a comparatively small perturbation can become critical for a path operating close to its timing requirement. The digital twin accounts for this distinction through $P_{\mathrm{fail}}$and the residual-slack calculation rather than through a universal delay threshold.

Across all application designs, 71.6% of the analyzed sink paths contained at least one valid adjacent dormant branch, while only 8.3% contained a candidate with $P_{\mathrm{fail}} \geq 0.50$. The highest-risk candidates were associated with shared routing trunks feeding low-slack endpoints and with serially extended dormant components. Several of these candidates did not belong to the longest-delay paths in the nominal implementation, demonstrating that conventional critical-path analysis alone does not identify all configuration-induced timing vulnerabilities.

### *H. Design-Level Vulnerability Analysis*

The design-level evaluation uses four independently implemented benchmarks: a five-stage in-order RV32I processor with separate instruction and data memories and execute-stage branch resolution; an eight-lane 32-bit Q16.16 multiply-accumulate datapath with a pipelined reduction tree and saturation logic; an AXI4-Stream-to-dual-port BRAM burst-buffer controller with address generation, width adaptation, and data-alignment logic; and a two-master/four-slave AXI4/AXI4-Lite subsystem with round-robin arbitration, register slices, and four programmable-logic accelerator endpoints. All benchmarks are implemented using the same XCZU7EV part, Vivado 2025.1 release, and post-route timing-analysis flow. Their source revisions, target frequencies, post-route resource utilization, routed-sink counts, enumerated candidate counts, and high-risk candidate fractions are summarized in Table III. For each design, the flow enumerated adjacent candidate PIPs, predicted the corresponding delay increments, associated the candidates with post-route slack, and generated the vulnerability score in Equation (38). The number of evaluated candidates scaled primarily with routed-net count and average inactive-PIP adjacency. The RISC-V core exhibited the highest fraction of candidates capable of exhausting path slack, primarily along tightly constrained control, forwarding, and pipeline-boundary paths. Ranking quality was evaluated from the overlap between the predicted and measured top-$k$ candidate sets. The model achieved a Recall@10 of 0.90 and a Recall@25 of 0.88, demonstrating its ability to identify the most consequential routing resources for targeted monitoring and selective protection.

Fig. 7 presents the spatial vulnerability atlas derived from the aggregated candidate-level scores over the implemented benchmark set. The left panel shows the raw atlas, obtained by accumulating the vulnerability contributions of all candidate perturbations mapped to each physical region of the XCZU7EV routing fabric. Indeed, Fig. 7(a) shows

This view highlights the absolute concentration of predicted

$$\tilde{A}_{sum}(x,y) = \frac{A_{sum}(x,y)}{\max_{x,y} A_{sum}(x,y)}$$

timing risk and therefore emphasizes the regions where both candidate density and predicted severity are jointly high. The right panel shows the corresponding density-normalized atlas, in which the accumulated score in each region is normalized by the number of enumerated candidates associated with that region. Similarly, Fig. 7(b) shows

This normalization suppresses the trivial effect of candidate

$$\tilde{A}_{mean}(x,y) = \frac{A_{mean}(x,y)}{\max_{x,y} A_{mean}(x,y)}$$

count and reveals whether a region remains intrinsically vulnerable after accounting for enumeration density. The persistence of the dominant hotspots in both panels indicates that the most critical regions are not merely those containing more candidates, but those whose routing context

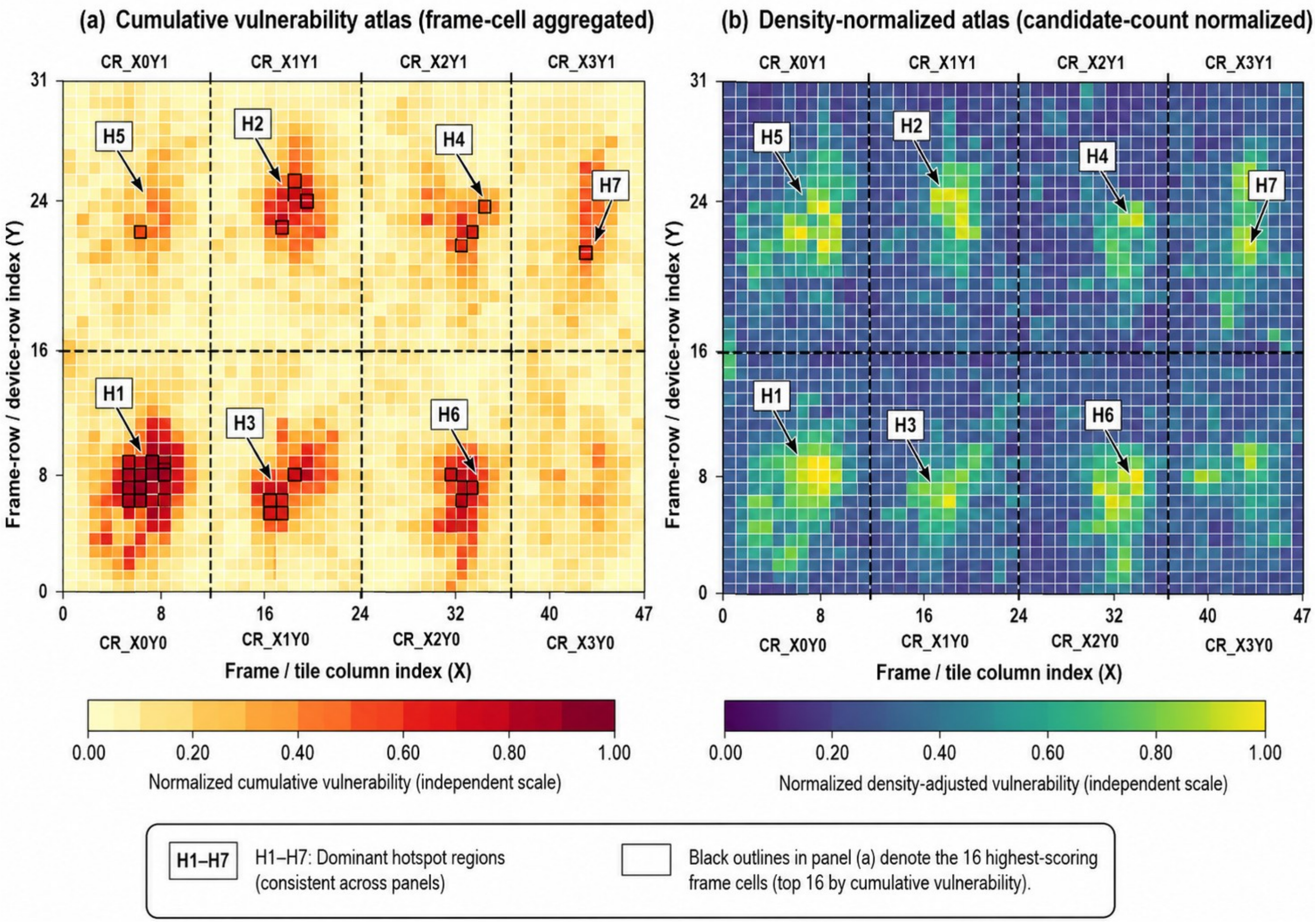

Fig. 7. Two-panel spatial vulnerability atlas for the implemented design on the XCZU7EV programmable logic fabric. **(a)** Raw aggregated vulnerability atlas, where each cell intensity represents the accumulated vulnerability score $A(x,y)$ over the corresponding spatial location. The outlined cells indicate the 16 configuration-frame locations with the highest predicted timing risk. **(b)** Density-normalized atlas, where the aggregated vulnerability is normalized by the number of enumerated candidates in each cell to distinguish intrinsic susceptibility from candidate-population effects. Dashed lines denote clock-region boundaries, and H1–H7 mark the principal hotspot regions that remain prominent across both views.

systematically produces larger delay perturbations and higher slack-exhaustion risk.

As shown in Fig. 7, the predicted timing vulnerability is distributed nonuniformly across the XCZU7EV programmable fabric. In the raw aggregated atlas of Fig. 7(a), a limited number of spatial regions account for a disproportionately large share of the total vulnerability score, indicating concentrated timing-risk hotspots rather than a uniform background distribution. These hotspots arise from the joint effect of branch topology, routing class, downstream loading, and residual slack. Fig. 7(b) presents the corresponding density-normalized atlas, in which the aggregated vulnerability is normalized by the number of enumerated candidates in each spatial cell. The persistence of the same dominant regions after normalization confirms that the observed hotspot structure is not explained solely by candidate density. Instead, these regions are intrinsically more susceptible to configuration-induced delay perturbations and therefore constitute natural targets for monitor placement, vulnerability-aware routing, prioritized scrubbing, and adaptive recovery policies.

### *I. Runtime and Scalability*

As already reported in Table II, the one-time device-graph extraction and model-fitting costs are practical for offline analysis, while millisecond-scale cumulative evaluation and submillisecond single-candidate prediction permit large routed designs to be screened without rerunning implementation.

### *J. Discussion and Limitations*

The results demonstrate that the delay effect of a configuration-induced routing attachment can be predicted more accurately by combining device-native graph information, a calibrated electrical representation, and path slack than by relying on topology, nominal STA, or configuration location alone. The proposed method also provides explicit uncertainty, allowing poorly supported routing classes or physical regions to be distinguished from well-calibrated predictions. The methodology does not claim recovery of the proprietary transistor-level FPGA interconnect. Its resistance and capacitance values are effective parameters identified from controlled hardware measurements. Their purpose is to reproduce the timing consequences of routing attachments while preserving structural interpretability.

The experiments reproduce configuration-equivalent attachment states deterministically. They validate the electrical delay mechanism previously observed under radiation in [5]-[7], but they do not independently determine particle cross sections, event rates, or mission-specific failure rates. When radiation-specific configuration-bit cross sections become available, they can be incorporated through the exposure term $\pi_B$without altering the delay-prediction framework. The reported model is specific to the locked XCZU7EV device and Vivado release. Transfer to another UltraScale+ part requires re-extraction of the device graph and partial electrical recalibration. The proposed class-based representation is intended to reduce that effort, but direct transfer accuracy must be established experimentally rather than assumed.

Finally, the current cumulative search considers structurally related perturbations rather than every possible combination of configuration changes. This restriction is necessary because unrestricted combination enumeration is exponential. The retained cases nevertheless cover the physically relevant parallel and serial attachments most likely to produce compounded loading on the same active route.

## VII. CONCLUSION AND FUTURE WORK

This paper presented a device-calibrated graph–circuit digital twin for predicting configuration-induced routing-delay degradation in Zynq UltraScale+ FPGAs. The proposed framework combines device-native routing-graph extraction, configuration association, an effective RC model, post-route slack analysis, and hardware calibration using routing-dominated ring oscillators and phase-swept path monitors. It therefore translates an unintentionally activated routing branch into a predicted delay increment, residual slack, uncertainty interval, and timing-failure probability.

For the representative XCZU7EV dataset, the model evaluated 864 routing perturbations and achieved a held-out MAE of 7.8 ps, an RMSE of 11.2 ps, and $R^2 = 0.908$. It also achieved an $F_1$score of 0.890 for slack-exhaustion classification and Recall@10 of 0.90 for identifying the most vulnerable candidates. The results further showed that cumulative perturbations, particularly serially extended branches, must be evaluated as unified graph–circuit structures rather than as simple sums of isolated PIP effects. The resulting PIP-, net-, frame-, and region-level vulnerability scores can support targeted monitoring, selective scrubbing, reliability-aware routing, and adaptive route recovery.

Beyond prediction accuracy, the proposed formulation provides an interpretable connection between configuration-level routing changes and their circuit-level timing consequences. By retaining the physical structure of the affected routing resources, the framework can distinguish vulnerability caused by local parasitic loading from that arising primarily from limited path slack. The spatial aggregation of candidate-level scores further enables device-level identification of concentrated vulnerability regions rather than treating configuration resources as uniformly critical. At the same time, the present results remain specific to the evaluated XCZU7EV implementation, routing database, and calibration conditions; therefore, absolute model parameters should not be assumed to transfer unchanged across FPGA families or implementation-tool releases. Such transfer requires device-specific routing extraction and, where necessary, recalibration of the electrical and residual model parameters.

Future work will extend the methodology to additional UltraScale+ devices, incorporate measured configuration-bit upset cross sections, and integrate the vulnerability score directly into placement and routing objectives. Further investigation will also address efficient multi-event analysis and experimental validation under representative radiation environments.

## ACKNOWLEDGMENT

The author would like to thank the École de technologie supérieure (ÉTS), Department of Electrical Engineering, and CMC Microsystems, Kingston, ON, Canada, for providing access to advanced design tools.

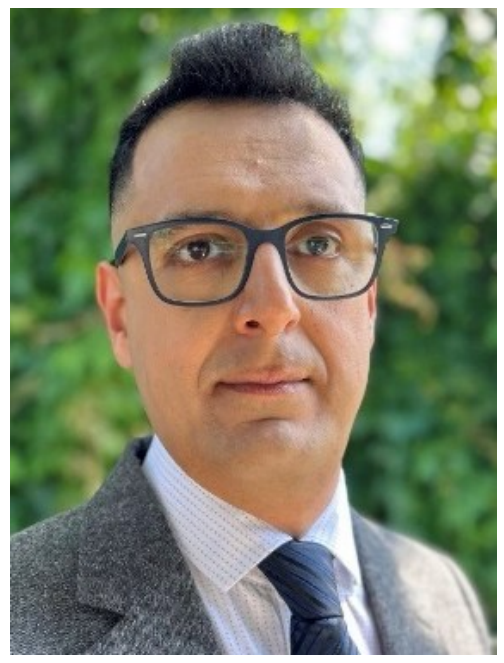

**Mostafa Darvishi** (Senior Member, IEEE) received the Ph.D. degree in Electrical Engineering from Polytechnique Montreal in 2018. He is currently with Electrical Engineering Department of École de technologie supérieure (ÉTS), Montreal, Canada. His research interest is mainly FPGA design for high-performance computing and embedded systems and radiation effects in avionic electronics and embedded systems.